# Investigating radioactivity in soil samples from neutral and vegetation land of Punjab/India

Sanjeet S. Kaintura[1*], Swati Thakur[1], Sarabjot Kaur[2], Soni Devi[1], Katyayni Tiwari[1], Priyanka[1], Arzoo Sharma[1], Pushpendra P. Singh[1, 2]

[1]Department of Physics, Indian Institute of Technology Ropar, Rupnagar – 140001, Punjab, India

[2]iHub – AWaDH, Indian Institute of Technology Ropar, Rupnagar – 140001, Punjab, India

*Email address: sanjeet.19phz0010@iitrpr.ac.in

**Abstract**

In this work, radioactivity investigations of soil samples from neutral and agricultural sites in Punjab/India have been carried out to study the impact of land use patterns. The analysis of radiological, mineralogical, physicochemical, and morphological attributes of soil samples has been performed employing state-of-the-art techniques. The mean activity concentration of $^{238}$U, $^{232}$Th, $^{40}$K, $^{235}$U, and $^{137}$Cs, measured using a carbon-loaded p-type HPGe detector, in neutral land was observed as 58.03, 83.95, 445.18, 2.83, and 1.16Bq kg$^{-1}$, respectively. However, in vegetation land, it was found to be 40.07, 64.68, 596.74, 2.26 and 2.11Bq kg$^{-1}$, respectively. In the detailed activity analysis, radium equivalent ($Ra_{eq}$) radioactivity is found to be in the safe prescribed limit of 370Bq kg$^{-1}$ for all investigated soil samples. However, the dosimetric investigations revealed that the outdoor absorbed gamma dose rate (96.08nGy h$^{-1}$) and consequent annual effective dose rate (0.12mSv y$^{-1}$) for neutral land, and the gamma dose rate (82.46nGy h$^{-1}$) and subsequent annual effective dose rate (0.10mSv y$^{-1}$) for vegetation land marginally exceeded the global average. The surface morphology of neutral land favored more compactness, while agricultural land favored high porosity. Various heavy metals of health concern, namely As, Cd, Co, Cr, Cu, Hg, Pb, Se, and Zn, were also evaluated in all soil samples using Inductively Coupled Plasma-Mass Spectroscopy (ICP-MS). Pollution Load Index (PLI) and Ecological Risk Index (RI) revealed that vegetation land was more anthropogenically contaminated than neutral land, with maximum contamination from Hg and As.

## Introduction

The fact that the inherent and ubiquitous natural radiations constitute an unavoidable feature of life on Earth (IAEA, 2003), and the naturally occurring radionuclide materials (NORMs) in terrestrial environments occupy the maximum share of radiation exposure to humans (UNSCEAR, 2000; WHO, 2009). The formation of these radionuclides ($^{238}$U, $^{232}$Th, $^{40}$K, and $^{235}$U) is due to the nucleosynthesis in stars having very long half-lives, ~$10^8$ to $10^{10}$ years. The worldwide average annual effective dose due to natural sources is estimated to be 2.4mSv but varies appreciably based on geographical locations, mineralogy, and individual characteristics (age, sex, etc.).

Natural radioactivity in a particular geographical region can be investigated by analyzing soil, water, air, and local vegetation samples. However, the concentration of radionuclides in the soil directly or indirectly dictates radioactivity in other environmental media (Durusoy & Yildirim, 2017). Soil is the predominant source and the medium for migrating radionuclides via groundwater sources or agriculture crop uptake. The external (cloud shine, ground shine, or contamination on skin or clothes) or internal exposure (inhalation, ingestion, or injection) to radionuclides may be potent for human health, leading to various stochastic effects, e.g., cancer incidence and cell mutation (Bramki et al., 2018). In addition, anthropogenic radionuclides such as $^{137}$Cs due to fallout from nuclear weapons or medical procedures can be detrimental in the long run (Michel, 1999). Thus, regular environmental radioactivity monitoring assists in understanding epidemiological radiation effects, maintaining nationwide baseline records, and checking against anthropogenic radiation spikes.

The soil itself is a complex composition of small particles having a size of a nanometer to millimeter level and plays a significant role in radionuclides distribution (UNSCEAR, 2000). The different factors such as the impact of climate, natural phenomena, sedimentation, leaching, dilution with varying minerals in various compositions, and percolation from groundwater with precipitation, etc., can influence the soil profile (physiochemical, morphological and mineralogical properties), which further alters the distribution of existing radionuclides (Dowdall and O'Dea, 2002; Navas et at., 2007). Physico-chemical soil inspection sheds light on migration (mobility and leaching), dispersion, sorption, and the biological impact of existing radionuclides in the terrestrial environment (Lee et al., 1997). Particle size, porous nature, and microstructures reveal the nature of the soil, its suitability for farming, and possible radionuclide distribution (Nenadović et al., 2012). Soil behaves as a source of heavy metals (HMs) and a sink to living organisms (humans and animals) via the food chain, surface, and ground waters (Facchinelli et al., 2001). Some HMs are essential to food crops; however, they induce toxicity above threshold limits. Heavy metal-laden food crops are a severe health concern as they do not decay quickly with time, unlike other radionuclides and organics (Facchinelli et al., 2001). HMs, as pollutants, have the potential to accumulate in different human organs and cause adverse health effects such as skin cancer, weakening of bones, hypertension, cardiovascular, neurological disorders, respiratory illness, etc., even at low concentrations (Sharma et al., 2018).

The Rupnagar (Ropar) region in the state of Punjab in India was submerged due to an overflow of water from the Sutlej River in 2019, which led to a considerable alteration in intrinsic soil properties. The present research work was conducted to analyze the impact of the flood on soil properties and compare the changes with that of an undisturbed (neutral) land representative of natural geology. Thus, the present study was intended to achieve the following objectives: (a) To compare the activity concentration of natural radionuclides ($^{238}$U/$^{226}$Ra, $^{232}$Th, $^{40}$K) and artificial radionuclides ($^{137}$Cs) in neutral and vegetation lands; (b) To estimate dose rate and annual effective dose to the members of the public; (c) To draw a contrast of physiochemical characteristics of soil (pH, electrical conductivity (EC) and porosity) and surface morphology of soil samples from both sites; (d) To statistically analyze correlations between radionuclides of interest and physicochemical parameters of soil; (e) To determine elemental composition and heavy metal concentration for pollution assessment.

**Experimental methods**

Study area

For quantitative determination of environmental radioactivity, a total of 20 soil aliquots (10 from each type of land use) were collected from neutral (A) and from vegetation land (B) in the Rupnagar district of Punjab, India. The geographical location of the neutral land extends from 30˚58′12″ to 30˚58′16″ N latitude and 76˚28′12″ to 76˚28′17″ E longitude, while for vegetation land, it is from 30˚58′24″ to 30˚58′28″ N latitude and 76˚32′21″ to 76˚32′24″ E longitude. Rupnagar is a major agricultural hub and receives one of the highest rainfall in Punjab (about 776 mm of rainfall annually). Ropar district is classified into four central lithological units: the Shivalik hills, the Kandi/Sirowal formation, the Sutlej intermontane valley, and the alluvial plain. The uneven topography of Shivalik hills separates the Himalayan ranges from the Indo-Gangetic alluvial plain. Besides the mountains, a 5 km inter-montane valley extends from Nangal to Ropar. Besides the valley, an alluvial fan marks the transitional region between the alluvial plains and Himalayan foothills. The coarse fraction of sediments brought down by hill torrents coalesce to form alluvial fan deposits. The soil texture in the district is usually loam or silty clay loam except along the Sutlej River and seasonal canals, where some sandy patches may be found. The Sutlej, Punjab's longest river, is a crucial irrigation source and flows near the area under investigation. Neutral Land was devoid of anthropogenic perturbations, such as using fertilizers or other impurities, for over 12 years. However, vegetation land was used extensively for cultivation, including many farm-based activities, such as plowing, irrigating, manuring, adding fertilizers, etc. All the sampling positions were carefully pointed using a handheld GPS (Global Positioning System), as shown in Figure 1.

Sampling and pre-processing of soil samples

The soil specimens were scraped 30cm below the ground using an auger tool to avoid surface contamination. About 1 kg of sub-surface, undisturbed soil sample was collected from each chosen site. Foreign impurities like pebbles, glass pieces, twigs, and debris were removed as they are irrelevant and may give erroneous analytical results. The soil samples were packed in tagged polythene bags for transport to the laboratory. They were air dried at room temperature for about 24 hours, followed by crushing with mortar and pestle to obtain a fine powder. The samples underwent heat treatment in an oven at 110˚C for 24 hours to remove moisture and obtain a constant mass. After cooling to room temperature, samples were sieved through a 150μm scientific sieve to get uniform grain quality. Post-sieving, they were packed into airtight plastic containers and sealed using adhesive tape. Finally, they were kept untouched for 6-8 weeks (>7 half-lives of $^{222}$Rn) before gamma spectroscopic analysis to attain secular equilibrium between $^{226}$Ra, $^{222}$Rn, and its daughters (IAEA, 1989). The utilized sample geometry (petri dish) is desirable as it completely covers the front portion of the detector, ensuring maximum efficiency.

Gamma-ray spectrometry set-up

IIT Ropar Low Background Measurement setup composed of a high resolution, coaxial p-type HPGe detector (model GEM30P4-83-RB, Ortec) of relative efficiency ~33% for characteristic energy (1332.5keV) of $^{60}$Co at 25cm. The detector has a 62mm diameter and 46mm length with a 0.9mm carbon fiber enclosure surrounded by passive lead shielding. The experimental setup (in a shielded configuration) is displayed in Figure 2. The spectrometer was operated at +2800 kV bias voltage. Electronic signals were amplified, pulse-shaped, and sorted using a multi-channel analyzer with 8K channels coupled with a data acquisition system. The acquired data were analyzed offline using Linux Advanced Multi-Parameter System (LAMPS: https://www.tifr.res.in/~pell/lamps.html) software. Before the sample measurement, the detector was calibrated for energy (from 121.7keV to 1408.1keV) with a $^{152}$Eu point gamma source. Efficiency calibration was performed by GEANT4 (for GEometry ANd Tracking) based simulation platform using Monte Carlo methods (Agostinelli et al., 2003) and verified by the known specific activity of $^{238}$U, $^{232}$Th, and $^{40}$K present in same soil sample, while to ascertain the efficiency of $^{137}$Cs interpolation method was used. The measured energy resolution (FWHM) was 0.79keV at 122keV and 1.73keV at 1332.5keV. The peak-to-Compton ratio using $^{60}$Co was 62:1. The Minimum Detectable Activity (MDA) was computed using the Curie equation (Curie, 1968):

$$\mathrm{MDA} = \frac{2.706+4.653\sigma_B}{P\times\varepsilon\times t} \quad (1)$$

Where $\sigma_B$ is the standard deviation in the background counts $N_B$, P is the transition probability of a particular gamma energy, ε is the absolute efficiency of the detector at the same energy, and t is the counting time per sample. The average MDA for $^{238}$U, $^{232}$Th, $^{40}$K and $^{137}$Cs were calculated as 2.44Bq $kg^{-1}$, 1.37Bq $kg^{-1}$, 10.30Bq $kg^{-1}$ and 0.03Bq $kg^{-1}$ respectively. Further, the detailed methodology for gamma spectroscopy utilizing a low-background HPGe detector with a carbon fiber window can be found in ref. (Thakur et al., 2024).

Measurement of specific activity, $A_X$, of radionuclides

To quantify the radioactivity concentration in soil samples, the specific activity is defined as the number of radioactive disintegrations per unit time per unit volume or activity per unit mass of the sample. For an individual radionuclide present in the soil samples, the specific activity ($A_X$) for a radionuclide 'X' is given by the following equation (UNSCEAR, 2000):

$$A_X(\mathrm{Bqkg^{-1}}) = \frac{C}{P\times t\times \mathcal{E}\times M} \quad (2)$$

Where C is net count rate, P is the transition probability, t is the lifetime of the measured sample (t = 86400s), M is the dry mass of the sample (in kg), ε is the absolute photopeak efficiency of the germanium detector at a particular energy. The transition probability of various gamma transitions of concerned radionuclides is tabulated in Table 1. The error propagation method ascertained the uncertainties, accounting for relative standard uncertainties of total counts in the peak, emission probability, full energy peak efficiency, and sample weight (Aközcan et al., 2018).

The natural abundance of $^{235}$U is very low (0.72%) due to its complex quantification. The energy line at 185.7keV is the most intense peak (57%) among the all-emitted gamma lines of $^{235}$U but is overlapped by another gamma line at 186.2keV due to the decay of $^{226}$Ra to $^{222}$Rn in the $^{238}$U chain with an intensity of 3.64%. Consequently, the resolution of the detector is questionable due to the overlapping of two characteristic gamma lines. This interference was removed by subtracting the contribution of $^{226}$Ra from that of $^{235}$U using the following relation (Powell et al., 2007; Wahl, 2010):

$$A_{235_U} = \frac{\left(\left(\frac{CR_{186}}{M\times\mathcal{E}}\right) - A_{226_{Ra}} \times P_{226_{Ra}}\right)}{P_{235_U}} \tag{3}$$

$CR_{186}$ is the count rate at the gamma line of energy 186keV, ε is the absolute efficiency of the detector at the energy of 186keV, M is the mass of the sample, $\mathbf{P}_{226_{Ra}}$ is the gamma emission probability for $^{226}$Ra at 186keV. $\mathbf{P}_{235_U}$ is the gamma emission probability for $^{235}$U at 186keV. $\mathbf{A}_{226_{Ra}}$ is the activity concentration of $^{226}$Ra in the soil sample (in Bq $kg^{-1}$).

Physico-chemical characterization

To measure the pH and electrical conductivity, 4gm soil was treated with 20mL of deionized water to obtain a mixture in a ratio of 1:5. The suspension was thoroughly shaken and filtered by 125-micron filter paper. Each filtrate was kept for 15 hours for uniform distribution of ions. A digital Oakton 700 pH/conductivity meter at room temperature deduced the soil pH and electrical conductivity. Porosity was measured by water-filled pore space or relative saturation method (Hao et al., 2008).

Scanning electron microscope and energy dispersive x-ray spectroscopy

The surface morphology of collected soil samples from both sites was examined by Scanning Electron Microscope (SEM) (JEOL JSM-6610 LV) working under ultra-high vacuum mode. SEM was operated at a voltage of 10kV as per the resolution requirement. Elemental composition in soil samples was diagnosed by the Energy Dispersive X-ray (EDX) spectrometer (Bruker XFlash 6130) equipped with SEM. The accelerating voltage was fixed at 20kV. Each soil specimen was scanned at two different regions to ensure homogeneity of observed elemental distribution.

Inductively coupled plasma mass spectroscopy

All the soil samples were digested using the method of aqua regia (Ehi-Eromosele et al., 2012). 0.1 grams of a fine soil sample was digested with 1 mL of concentrated $HNO_3$, 2 mL of HF, and 3 mL of HCl on a microwave digestor until complete digestion. The samples were cooled, filtered, and diluted correctly using deionized water to get a final volume of 100 mL. Lastly, they were stored in glass containers for heavy metal analysis.
The digested samples were examined for trace elemental concentration by quadrupole system-based Inductively Coupled Plasma-Mass Spectrometer (ICP-MS). ICP-MS (Agilent, model-8900 ICP-MS Triple Quad) provides reliable detection of essential concentration in different samples with a detection limit of up to sub parts per billion (ppb). The instrument was calibrated before measurement using standard solutions of different concentrations to obtain precise calibration factors. The soil samples of interest were ionized with an inductively coupled plasma source, and then mass spectrometry was employed to detect and quantify various elements.

## Results and discussion

Radiological investigations

The radiological investigation provides knowledge about background radiation levels in a specific geographical region and possible health impacts on surviving humans in that environment due to exposure to these radiations. Measuring the gamma-emitting radioactivity in the soil sample represents the background radiation level in that region. There are various radiometric methods and techniques to assess the radioactivity present in the soil but the gamma spectroscopy technique using HPGe in the laboratory is a non-destructive technique and the most suitable method for accurate and precise measurement (Trang et al., 2021; IAEA, 2003).

Assessment of radionuclide concentrations

The native soil samples from sites A and B were assessed to quantify the level of terrestrial radioactivity by employing the high-resolution HPGe spectrometer. Figure 3 confirms the suitability of the experimental setup for monitoring environmental samples with radioactivity above the ambient background. The observed activity concentration of radionuclides of interest, $Ra_{eq,}$ and calculated doses, along with the uranium isotopic ratio for each soil sample, is summarized in Table 2. In neutral land (A), the activity concentration of $^{238}U$, $^{232}Th$, and $^{40}K$ varied from 37.00 ± 1.17 to 76.13 ± 1.96 Bq $kg^{-1}$, 55.58 ± 1.43 to 106.82 ± 2.53 Bq $kg^{-1}$, 381.37 ± 23.49 to 526.26 ± 31.69 Bq $kg^{-1}$ respectively. In vegetation land (B), the activity concentration of $^{238}U$, $^{232}Th$, and $^{40}K$ varied from 34.06 ± 1.03 to 44.79 ± 1.28 Bq $kg^{-1}$, 56.94 ± 1.49 to 70.04 ± 1.75 Bq $kg^{-1}$, 438.11 ± 26.00 to 712.90 ± 42.02 Bq $kg^{-1}$ respectively. The $^{238}U$, $^{232}Th$, and $^{40}K$ activity concentrations at both sites are displayed as bar graphs in Figures 4(a) and 4(b). A higher average concentration of $^{238}U$, $^{232}Th$, and $^{235}U$ was observed in the case of neutral land as compared to vegetation land, which may be due to the intrinsic mineralogy of soil. However, the results were the opposite regarding average activity concentration of $^{40}K$ and $^{137}Cs$. A greater activity concentration of $^{40}K$ in vegetation land may be attributed to the indiscriminate spraying of fertilizers like NPK (Nitrogen, Phosphorus, and Potassium) and high clay content. $^{40}K$ is a constituent of clay minerals, and its mobility is controlled by solubility in the soil (Bajoga et al., 2017). The average activity concentration of $^{137}Cs$ in neutral and vegetation land was 1.16Bq $kg^{-1}$ and 2.11Bq $kg^{-1}$, respectively. Highly reactive $^{137}Cs$ behave as cations like $K^+$, $NH_4^+$, or $H^+$, as well as minerals $Na^+$, $Ca^+$, or $Mg^+$. Hence, the vertical transportation (as samples were collected 30cm below the surface) of $^{137}Cs$ was faster in vegetation land as it has fine, conducting soil particles compared to neutral land with large coarse particles in soil (Ahmad et al., 2019). At both sampling sites, A and B, the order of radioactivity levels followed the trend – $^{40}K > ^{232}Th > ^{238}U > ^{235}U > ^{137}Cs$. This confirms that $^{40}K$ is the most dominant gamma radioactivity source in terrestrial environs and is a usual primary weathering product (Guagliardi et al., 2016).

The average activity concentration of various natural radionuclides in surveyed regions was also compared with that in the entire state of Punjab, India, and the world. The average activity concentration of $^{238}U$ at neutral and vegetation land was less than that reported for Punjab (55.50Bq $kg^{-1}$) (Saini & Bajwa, 2017). This may be accredited to the easy migration of $^{238}U$ in the absorbed form of iron hydroxides, clay particles, and suspended organic matter via flood water at IIT Ropar in 2019 (Iskra & Bakhurov, 1981). The hexavalent uranyl ion also forms stable complex compounds on the surface of $SiO_2$ (Dent et al., 1992). The average activity concentration of $^{232}Th$ at both neutral land (83.95Bq $kg^{-1}$) and vegetation land (64.68Bq $kg^{-1}$) significantly surpasses the state and national averages of 21.00Bq $kg^{-1}$ (Saini & Bajwa, 2017) and 30.96Bq $kg^{-1}$ (Ramachandran, 2011) respectively—moreover, these concentrations elevated beyond the global average (UNSCEAR, 2008). For $^{40}K$, the mean activity concentration values from neutral land were consistent with state (443.40Bq $kg^{-1}$), national (432.70Bq $kg^{-1}$), and global values (420Bq kg-1), but for vegetation land, it exceeded all these values. The activity concentration of $^{137}Cs$ was appreciably less than the national (3Bq $kg^{-1}$) (Kumar et al., 2012) and global values (51Bq $kg^{-1}$) (UNSCEAR, 2008).

The isotopic ratio of $^{235}U/^{238}U$ ranged from 0.041-0.060 in neutral land and 0.051-0.060 in vegetation land. The dominant contribution to natural radioactivity was observed from $^{238}U$, $^{232}Th$, and $^{40}K$, along with a minor contribution from $^{235}U$ and trace amounts from $^{137}Cs$.

Radium equivalent activity ($Ra_{eq}$)

Radium equivalent activity ($Ra_{eq}$) gives the equivalent therapeutic effect of the activity of different radionuclides as that of 1 mg of $^{226}Ra$ (Tufail, 2012). It assumes that the activity concentration of 370Bq $kg^{-1}$ of $^{226}Ra$ or 259Bq $kg^{-1}$ of $^{232}Th$, or 4810Bq $kg^{-1}$ of $^{40}K$ spread uniformly produces an outdoor external effective dose rate of 1.5mGy (1mSv) per annum at 1m distance from the material (Krisiuk et at., 1971). Since there is a non-uniform distribution of $^{226}Ra$, $^{232}Th$, and $^{40}K$ in soil or rocks, the cumulative gamma output of these three significant primordial radionuclides is often expressed in terms of radium equivalent, as given by (NEA, 1979):

$$\mathrm{Ra_{eq}} = \left(\frac{A_{Ra}}{370} + \frac{A_{Th}}{259} + \frac{A_K}{4180}\right) \times 370 \tag{4}$$

Where $A_{Ra}$, $A_{Th}$, and $A_K$ are the activity concentrations of $^{226}Ra$, $^{232}Th$, and $^{40}K$, respectively.

The spread in $Ra_{eq}$ values in soil samples belonging to neutral land was more than in vegetation land. Thus, the distribution of $^{238}U$, $^{232}Th$, and $^{40}K$ was more uniform in vegetation land, which is expected due to various farm activities like plowing, leveling, etc. $Ra_{eq}$ varied from 153.39 to 258.96 Bq $kg^{-1}$ in neutral land and from 150.31 to 194.53 Bq $kg^{-1}$ in vegetation land with arithmetic mean values of 212.20Bq $kg^{-1}$ and 178.37Bq $kg^{-1,}$ respectively. The $Ra_{eq}$ value of all samples was well within the advocated threshold of 370Bq $kg^{-1}$ given by the Organization of Economic Cooperation and Development (NEA, 1979). Also, the average $Ra_{eq}$ value in the surveyed region was comparable to the average value reported for the state of Punjab (Saini & Bajwa, 2017) but higher than that of India (106.11Bq $kg^{-1}$) (Ramachandran, 2011) and the world (108.6Bq $kg^{-1}$) (UNSCEAR, 2008).

Absorbed dose rate ($\dot{D}$)

The outdoor absorbed dose rate (in nGy $h^{-1}$) in the air from the experimentally measured value of natural radionuclides ($^{226}Ra$, $^{232}Th$, and $^{40}K$) at 1 meter above the earth's surface can be estimated using the following semi-empirical formula (UNSCEAR, 2000):

$$\dot{D}(\mathrm{nGy\ h^{-1}}) = 0.462\ A_{Ra} + 0.604\ A_{Th} + 0.0417\ A_K \tag{5}$$

Here, the contributions of other terrestrial radionuclides like $^{137}Cs$ and $^{90}Sr$ should have been addressed. For neutral land, the outdoor air absorbed dose rate range was calculated as 70.72 - 116.08 nGy $h^{-1}$ (mean = 96.08nGy $h^{-1}$), while for vegetation land, it was 68.96 - 90.42 nGy $h^{-1}$ (mean = 82.46nGy $h^{-1}$). Figure 5 visually portrays the distribution pattern of air-absorbed dose ($\dot{D}$) through a contour map featuring multiple contour lines. Therefore, all samples from both site A and site B exhibited an absorbed dose rate higher than the global average of 58nGy $h^{-1}$.

Annual effective dose ($A_D$)

Annual effective dose is a scale to measure the health risk effects like deterministic and stochastic effects after irradiation due to a radioactive source. The yearly average effective dose in outdoor environments is calculated from the air-absorbed dose rate using the following relation (UNSCEAR, 2008):

$$A_D(\mathrm{mSv}) = \dot{D}(\mathrm{nGy\ h^{-1}}) \times 8760\ \mathrm{h} \times 0.2 \times 0.7(\mathrm{Sv\ Gy^{-1}}) \times 10^{-6} \qquad (6)$$

Where 0.7Sv $Gy^{-1}$ is the conversion factor to convert the absorbed dose rate (in nGy $h^{-1}$) to the equivalent effective dose rate (in mSv $y^{-1}$). 0.2 represents the outdoor occupancy factor. The recommended value of the annual effective dose is 0.07mSv $y^{-1}$ (UNSCEAR, 2008). The International Commission on Radiological Protection in 1990 recommended dose constraints of order 0.3 - 1 mSv per annum to the general public resulting from chronic exposure to natural terrestrial radiations (ICRP, 1991). Figure 6 depicts the visual representation of the distribution pattern of the Annual Effective Dose ($A_D$) using a contour map having multiple contour lines. The mean annual effective dose reported in neutral land (0.12mSv $y^{-1}$) and that in vegetation land (0.10mSv $y^{-1}$) was also found to be higher than the world mean value of 0.07mSv $y^{-1}$. Thus, the studied region has moderate background radiation levels. Consequently, there exists a possibility of stochastic health effects in humans when exposed to these low doses and low dose rates.

Physico-chemical assessment

It has been reported that the physiochemical properties influence radionuclide speciation, distribution, and migration (Tsai et al., 2011). These properties also hint at the soil's salinity and ion exchange capacity. Testing soil pH and conductivity is also an important prerequisite for selecting vegetation (Smith and Doran, 1997). Different precipitates, such as carbonates, complex phosphate, hydroxyl, or sulfide ions, are formed in an alkaline medium. These insoluble components in soil lower the presence of radionuclides on the upper soil surface. On the other hand, the ionic medium promotes the replacement of cationic radionuclides like $^{137}Cs$ with $H^+$ ions (Mesrar et al., 2017; Ahmad et al., 2019). The physiochemical parameters, namely, pH, electrical conductivity, and porosity of soil samples from both sites, are stated in Table 3. The soil samples from neutral land were more alkaline, with pH ranging from 8.16 to 9.07 with an arithmetic mean of 8.83 ± 0.26. The pH of alkaline soil samples from vegetation land varied from 8.34 to 8.85 with an arithmetic mean of 8.64 ± 0.15. Wheat, rice, maize, and mustard seeds are prime food crops grown in Ropar, Punjab. Around 95% of the state of Punjab is a hotspot for food crops in India, with the pH of the soil in the range of 6.5 to 8.5 (40% of the area has a pH range of 6.5 to 7.5, and 54% has 7.5 to 8.5). Hence, the present study did not show substantial changes in soil pH concerning previous studies conducted by Sharma et al. (2016).

The electrical conductivity (EC) of soil is a significant indicator of salinity present in the soil. The range of EC of soil for neutral and vegetation land was found to be 46.26 - 74.17 µS $cm^{-1}$ and 73.73 - 145.30 µS $cm^{-1}$, respectively. The average conductivity of neutral land (58.02 ± 9.80 µS $cm^{-1}$) lower than that of vegetation land (103.92 ± 23.39 µS $cm^{-1}$) is expected due to the contribution from major ions like $K^+$ and $NO_3^-$. All these ions accumulate in the terrestrial matrix because of evaporation, transpiration, and high dosages of fertilizers (Visconti and de Paz, 2016). Soils with a higher content of smaller soil particles (higher content of clay) are more conducting than soils with a higher content of larger silt and sand particles (lower content of clay).

Porosity is another useful indicator to predict aeration status, water storage capacity, rate of water infiltration, class of soil structure, and water affordability to plants. The porosity of sandy surface soil typically ranges from 35% to 50%, while, for finer textured soil, it ranges from 40% to 60% (Hao et al., 2008). The mean porosity for site A was computed as 39.75 ± 2.49%, while that for site B was 44.99 ± 3.88%. The highly porous nature of vegetation land ensures quicker seepage, uptake by plants, and migration of radionuclides.

Pearson's correlation between radioactivity level and physicochemical parameters

The correlation between the activity concentration of radionuclides ($^{238}$U, $^{232}$Th, $^{40}$K, and $^{137}$Cs) and physicochemical properties (pH, EC, and porosity) for neutral and vegetation land was carried out using ORIGIN PRO 2023b and presented in Figure 7(a) and 7(b) respectively.

A strong positive correlation ($p \leq 0.01$) between activity concentrations of $^{238}$U and $^{232}$Th at both neutral (r=0.997) and vegetation land (r = 0.932) was observed, which is consistent with previous studies (Kannan et al., 2002; Mubarak et al. 2017). U and Th were also strongly correlated with K in neutral land with r = -0.815 for U v/s K and r = -0.808 for Th v/s K, while in vegetation, it had moderate relation ($p \leq 0.05$) with r = 0.661 for U v/s K and r = 0.769 for Th v/s K. Strong correlation implies that in this region $^{238}$U, $^{232}$Th and $^{40}$K have a same geogenic origin (mineral components of soil). A positive correlation of $^{137}$Cs was obtained with the $^{238}$U, $^{232}$Th, and $^{40}$K with the Pearson correlation values (r) of 0.823, 0.834, and 0.528, respectively, in vegetation land, which may be due to some artificial agricultural inputs. A negligible correlation of $^{137}$Cs with U, Th, and K was observed in neutral land, which may be due to negligible anthropogenic inferences and hence, the poor statistics of $^{137}$Cs.

The present study did not find any strong dependence of $^{238}$U, $^{232}$Th, and $^{40}$K on the pH of soil samples from both investigated sites. Since soil pH governs $^{137}$Cs radionuclide transportation, a negative correlation of $^{137}$Cs with pH was observed in both cases. A lower pH value supports the presence of $^{137}$Cs, which was reflected in our results (Ahmad et al., 2019). Since soil samples in both sites were collected 30cm below the top surface. Hence, a higher content of $^{137}$Cs was found in vegetation land, showing the rapid transportation of $^{137}$Cs in disturbed soil compared to undisturbed soil of Neutral land. A negative correlation was found between radionuclides ($^{238}$U, $^{232}$Th, $^{137}$Cs) with porosity, while a positive correlation was found between $^{40}$K and porosity in both sites. However, a strong correlation ($p \leq 0.001$) was observed for the $^{40}$K v/s porosity (r = 0.873), while a moderate correlation ($p \leq 0.01$) found $^{238}$U v/s porosity (r = -0.629) and $^{232}$Th v/s porosity (r = -0.629) in neutral land. Hence, the radioactivity present in the soil is affected by the porosity, indicating that the distribution of the radionuclides is influenced and conducted by the particle size or grain size distribution (Barisic et al., 1998; Ravisankar et al., 2019). However stronger influence of porosity was observed in neutral land. Also, a negative correlation was observed between pH, $^{238}$U with EC in both sites, while it was higher in vegetation land with a significant level of $p \leq 0.05$. No significant correlation was observed with EC. Hence physicochemical properties can be regarded as one of the significant factors in exploring the radioactive contents of soil (Ravisankar et al., 2019).

Surface morphology

Soil morphology enables an understanding of topography that governs soil's hydrological, physical, chemical, and micromorphological traits. The neutral land comprises a high quantity of sand, making it uniformly graded with low porosity. On the other hand, the vegetation land comprised ample silt and clay (along with traces of sand), making it well-graded clayey-silt soil with high porosity. A collection of soil particles in Figure 8(a) shows the uniform distribution of sandy soil grains on neutral land. Figure 8(b) represents the morphology of each type of soil particle in neutral land, which is less porous and more compact. Figure 8(c) zooms in on the same particle to showcase the clear structure of a single soil particle. Figure 9(a) shows the agglomeration of different soil particles in vegetation land having their submissive sizes. Figure 9(b) illustrates a single soil particle in agricultural land, which is less compact and more porous in structure. Figure 9(c) shows a zoomed portion of the same soil grain to get a better insight into its morphology. SEM analysis concluded that the morphology of vegetation land was strongly influenced by efflorescence due to ample clayey content. The exchange capacity of the soil coarse fraction may be responsible for the irregular angular shape of sand particles in neutral land (Nenadović et al., 2012).

Often, it has been observed that a finer fraction of grain size of soil and sediments results in higher activity concentrations of $^{238}$U, $^{232}$Th, and $^{40}$K (Narayana and Rajashekara, 2010). The increment in activity concentration of naturally occurring radionuclides in vegetation land is mainly due to the adsorption of radionuclides onto crystals, their grain boundaries, or the crystal defects (Baeza et al., 1995). So, for a specific core soil sample, a smaller grain size fraction with a larger surface-to-volume ratio results in greater adsorption of radionuclides, which supports the present study's activity concentration of $^{40}$K. However, the activity concentration of $^{238}$U and $^{232}$Th in soil from neutral land (having sandy nature) was a bit higher than that in vegetation land (composed of clayey silt), which might be due to the inherent mineralogy of soil content. Physio-chemical and mineralogical changes in the distribution of radionuclides occur due to physiographic alterations, weathering impact, and natural disasters (like flooding) in the region (Nenadović et al., 2012; Yasmin et al., 2018).

## Elemental composition

The elemental composition of major and minor elements in neutral and vegetation land soil was ascertained using EDX. EDX characterization of soil samples revealed that the major elements of soil were silicon (Si) and oxygen (O), which hinted at the presence of a high amount of $SiO_2$ (silica) in the soil at both probed sites. K content was higher in vegetation land owing to the indiscriminate use of fertilizers like NPK (Nitrogen, Phosphorus, and Potassium) for the cultivation of crops (ENVIS Centre, Punjab State Council for Science and Technology, Chandigarh: http://punenvis.nic.in/index3.aspx?sslid=5862&subsublinkid=4973&langid=1&mid=1). The average elemental composition for Al, Mg, Ca, and Fe in soils was found to be 8.93 ± 3.09%, 0.97 ± 0.14%, 4.91 ± 1.70%, 3.19 ± 0.41%, respectively, for neutral land and 11.62 ± 3.57%, 1.36 ± 0.33%, 3.16 ± 0.73% and 4.64 ± 0.37%, for vegetation land as depicted in Figure 10.

## Heavy metal Assessment

In a few years, the level of HMs in the environment has been enhanced beyond permissible limits due to various artificial inputs such as industrial and agronomic activities. A comprehensive study of HM concentration in representative soil samples is crucial to identifying, monitoring, and investigating the potential sources of contamination in the geographical region of interest (Nugraha et al., 2022).

### HM Concentration

A highly sensitive ICPMS technique was employed to establish the heavy metals (HMs) concentration of Chromium (Cr), Cobalt (Co), Copper (Cu), Zinc (Zn), Arsenic (As), Selenium (Se), Cadmium (Cd), Lead (Pb) and Mercury (Hg) in soil samples. The order of concentration of HMs in the neutral land was Cr > Zn > Pb > Cu > As > Co > Hg > Se > Cd with their respective mean values as 43.84, 42.56, 28.02, 11.32, 7.48, 5.83, 2.77, 1.86 and 0.24 ppm while in the vegetation land, Hg jumped three places ahead of Cu due to its higher concentration, setting the trend as Cr > Zn > Pb

> Hg > Cu > As > Co > Se > Cd with their means as 65.02, 58.00, 27.58, 19.51, 17.74, 9.58, 9.08, 2.61 and 0.31 ppm respectively. The mean concentration of all the HMs of interest in the vegetation land except Pb was higher than that in the neutral land. The higher concentration of heavy metals in the vegetation land might be due to the propinquity of sampling locations to the National Highway 21, intensive use of fertilizers, manure and pesticides, wastewater irrigation, and aggregation of residues during rainfalls (Gupta et al., 2021).

The statistical variation of heavy metal content in soil samples from neutral and vegetation land is described in Figures 11 (a) and 11 (b), respectively, with the help of Box and Whisker plots. The width of the box is decided by the 25$^{th}$ and 75$^{th}$ percentiles. The horizontal line separating the box into two compartments depicts the median value, and the symbol ‘□’ depicts the arithmetic mean value. The markers above and below the box indicate the 1$^{st}$ and 99$^{th}$ percentile, respectively. The prime statistical indicators of the HMs in the soil of both lands, along with their background values for Indian natural soil and permissible limits provided by different agencies, have been explicitly mentioned in Table 4. Among all heavy metals, Hg in the soil of vegetation land had an alarming toxicity level as its concentration surpassed the permissible limits set by WHO/FAO (2007) and USEPA (2002) of 0.3ppm and 1ppm, respectively. In neutral land, contamination due to mercury was obtained in the range of 1.44ppm to 4.78ppm with a mean value of 2.77 ± 1.25 ppm, while in vegetation land, it ranges from 6.87ppm to 46.97ppm with an average of 19.51 ± 15.94 ppm. Notably, the highest concentration difference at the two sites was observed in the case of Hg in vegetation land, which was about seven times higher than in neutral land. The mean concentration of Zn, Cd, and Pb in vegetation land was found slightly greater than the recommended limit by WHO/FAO, while, in neutral land, the mean concentration of only Pb crossed the advocated level. Nonetheless, the concentration of Co, Cu, Zn, Cd, and Pb in the soil samples was safely below the permissible Indian standard range as recommended by Awashthi (2000).

Pollution indices

Pollution indices hint at the overall adulteration of soil resulting from natural disasters or anthropogenic perturbations. Some widely used pollution indices are the contamination factor, pollution loaded index, geoaccumulation index, and ecological risk index. The soil is classified into various categories based on the value of pollution indices, as outlined in Table 5.

1. Contamination Factor (CF)

CF highlights the additional contamination and ecological hazard promoted by the existing HMs in the soil above the background. It is calculated by Hakanson’s formula (1980), which is simply the ratio of the concentration of HM present in investigated soil ($C_i$) to that in the natural background environment ($C_b$):

$$CF = \left(\frac{C_i}{C_b}\right) \quad (7)$$

The background value for Hg was taken from Wang et al. (2012) while for the rest of the HMs, they were taken from Taylor & McLennan (1995). The range and average values of all pollution indices for neutral and vegetation land have been summarized in Table 6.

CF for Co, Cu, Zn, and Se in all soil samples was less than unity, indicating low contamination due to these HMs. CF for Cr, Cd, and Pb in all soil samples had moderate contamination (CF < 3) and thus, require careful and regular monitoring. The maximum contamination in soil samples was due to As and Hg (CF > 6). The mean values of CF of As and Hg in neutral land were 4.99 and 27.68 respectively. The average value of CF of As and Hg further increased in vegetation land to 6.38 and 195.08 respectively, which may be attributed to irrigation water or flood water contaminated with these heavy metals.

2. Pollution Loaded Index (PLI)

The total level of contamination due to all the HMs of interest in each soil sample can be evaluated comprehensively by the Pollution Loaded Index (PLI). PLI was proposed by Tomlinson et al. (1980) and is given by:

$$\mathrm{PLI} = (\mathrm{CF}_1 \times \mathrm{CF}_2 \times \mathrm{CF}_3 \times \dots \dots . \times \mathrm{CF}_n)^{\frac{1}{n}} \quad (8)$$

Where n signifies the total number of metals of interest in each sample, the range of PLI values in neutral and vegetation land was calculated as 0.95-1.35 (1.12 ± 0.13) and 0.91-2.20 (1.76 ± 0.42). 83% of samples had PLI values greater than unity, signifying extreme deterioration of soil quality due to heavy metals. Figure 12 clearly shows that heavy metal pollution is more prominent in vegetation land than in neutral land.

3. Geoaccumulation Index ($I_{geo}$)

The Geoaccumulation index is broadly used to quantify the level of HM contamination in cultivated or urban land soil. Müller proposed this index (Igeo) in 1969. The following equation defines it:

$$\mathrm{I}_{\mathrm{geo}} = \log_2\left(\frac{\mathrm{C}_i}{1.5 \times \mathrm{C}_b}\right) \quad (9)$$

The constant (1.5) allows us to minimize the impact of possible variation in the background and to sense the low-level anthropogenic interferences. $I_{geo}$ values for Cr, Co, Cu, Zn, Se, Cd, and Pb in soil samples representing neutral and vegetation land fell in the category of "Uncontaminated to moderately contaminated". The average $I_{geo}$ value for As in neutral land was 1.00 ± 0.32, while in vegetation land increased to 1.28 ± 0.41, signifying moderate contamination. Furthermore, the maximum value of $I_{geo}$ values was observed for Hg among all heavy metals for both neutral (5.55 ± 2.49) and vegetation land (39.15 ± 31.98). The scrutinized region was highly contaminated with Hg ($I_{geo} \geq 5$).

4. Ecological Risk Index (RI)

The risk index is applied to determine the potential ecological risk factor in contaminated soil using the relative concentration of heavy metals in a sample concerning the background concentration. It indicates the toxicity of a biological substance and illustrates the potential ecological risk due to HM contamination (Nugraha et al., 2022). Mathematically, it is evaluated as (Hakanson et al., 1980):

$$\mathrm{RI} = \sum \mathrm{E}_i \quad (10)$$

$$\mathrm{E}_i = \mathrm{T}_i . \mathrm{f}_i = \mathrm{T}_r . \frac{\mathrm{C}_i}{\mathrm{C}_b} \quad (11)$$

$T_r$ is the toxic response factor of HM, $f_i$ is the ratio of individual HM concentration ($C_i$) to the background concentration ($C_b$) of HM present in soil and $E_i$ is the risk factor of an individual HM. The $T_r$ values for studied HMs in decreasing order are taken as Hg = 40 > Cd = 30 > As = 10 > Pb = Cu = 5 > Cr = 2 > Zn = 1 (Hakanson et al., 1980).

$E_i$ value for Cr, Cu, Zn and Pb was < 40 indicating low contamination. $E_i$ for As fell in the range of 40-80, suggesting that soil was substantially contaminated in both neutral and vegetation land. Hg warrants the maximum attention as $E_i$ value for Hg in both sites was greater than 320, signifying very high contamination of soil samples ($E_i > 320$). The range of $E_i$ values for Hg in neutral and vegetation land was estimated to be 574.00 - 1910.40 (1107.00 ± 498.10) and 1092.80 – 18786.40 (7803.33 ± 6375.04) respectively. Vegetation land was additionally contaminated with Cd with an average $E_i$ = 93.88.

The cumulative contamination of soil samples due to the presence of heavy metals was expressed in terms of the ecological risk index. RI value for soil in both neutral and vegetation land was found in the highest risk grade (RI > 600). The average RI computed for neutral land was 1243.53 ± 473.60, while that for vegetation land was 7976.03 ± 6389.69. Such life-threatening RI values are mainly due to high Hg content in soil samples.

This study highlights the high level of artificial contamination of agricultural farms with heavy metals as compared to neutral land. The soil burdened with mercury may cause deleterious health effects on the human population. Once absorbed by plants, mercury gets transported, bio-magnified, and accumulated in different internal human organs via the food chain. Heavy mercury intake can damage kidneys and the immune system and cause various neurodegenerative diseases like Parkinson's disease, Alzheimer's disease, and Amyotrophic lateral sclerosis (Mutter et al., 2004).

## Conclusions

The radioactivity concentration of $^{238}U$, $^{232}Th$, $^{235}U$, and $^{232}K$ present in soil samples from neutral and vegetation land was determined by a high-resolution HPGe detector. The mean activity concentration of $^{238}U$ and $^{232}Th$ was found to be higher in neutral land soil samples as compared to those analyzed from vegetation land. However, the activity concentration of $^{40}K$ was higher in vegetation land owing to the use of artificial fertilizers like NPK (Nitrogen, Phosphorus, and Potassium). The average annual effective dose was found to be higher in the case of neutral (0.12mSv $y^{-1}$) and vegetation (0.10mSv $y^{-1}$) land. The regions were further explored for physicochemical parameters and heavy metal content of soil samples to understand the impact of cultivation on the mineralogical characteristics of the underlying soil. Soil belonging to neutral land was a more alkaline, conductive, uniformly graded coarse fraction with less porosity than the spongy soil of vegetation land. Based on EDX analysis, it was observed that Si and O had high percentage concentrations indicating the presence of silica. Vegetation land was overloaded with heavy metals, namely Zn, Cd, Pb, and Hg, surpassing the advocated limits by WHO. Consequently, pollution indices in vegetation land have a much higher value than in neutral land.

## Acknowledgements

The authors thank the iHub – AWaDH, a Technology Innovation Hub established by the Ministry of Science & Technology, Government of India, at the Indian Institute of Technology Ropar in the framework of the National Mission on Interdisciplinary Cyber-Physical Systems (NM—ICPS). Dr. Anil Kumar Gourishetty and Dr. Ashish Kumar are acknowledged for providing the ICP-MS facility at the Indian Institute of Technology Roorkee; Dr. Resmi Sebastian, Department of Civil Engineering of IIT Ropar, for providing the sample preparation facility; Dr. S. Manigandan and Mr. Chandrashekhar, Department of Chemical Engineering of IIT Ropar for providing pH, EC, and porosity test facilities. One of the authors, Sanjeet. S. Kaintura thanks the Ministry of Education, Government of India for the doctoral fellowship.

## Authors contribution

Sanjeet S. Kaintura: conceptualization, sample collection, and preparation, experimental measurements, software, analysis, writing – preparation, editing, and reviewing the manuscript; Swati Thakur: conceptualization, validation, writing - editing and reviewing the manuscript; Sarabjot Kaur: validation, writing - preparation, editing and reviewing the manuscript; Soni Devi: sample preparation, experimental measurements, and data analysis; Katyayni Tiwari: sample preparation, experimental measurements, and data analysis; Priyanka: sample preparation, experimental measurements, and data analysis; Arzoo Sharma: sample preparation, experimental measurements, and data analysis; Pushpendra P. Singh: supervision, writing - editing and reviewing the manuscript.

**Funding**

This work is supported by the Ministry of Science & Technology, Government of India, through the Technology Innovation, AWaDH, established at the Indian Institute of Technology Ropar in the framework of the National Mission on Interdisciplinary Cyber-Physical Systems (NM—ICPS).

**Data availability**

All datasets used or analyzed during the study are available from the corresponding author upon reasonable request.

**Declarations**

**Ethical approval**

The authors have unanimously approved the submission of this paper.

**Consent to participate**

All authors consent to participate in this publication.

**Consent for publication**

All the authors agree to consent to the publication.

**Conflict of interest**

The authors declare that they have no competing interests.

**Figures:**

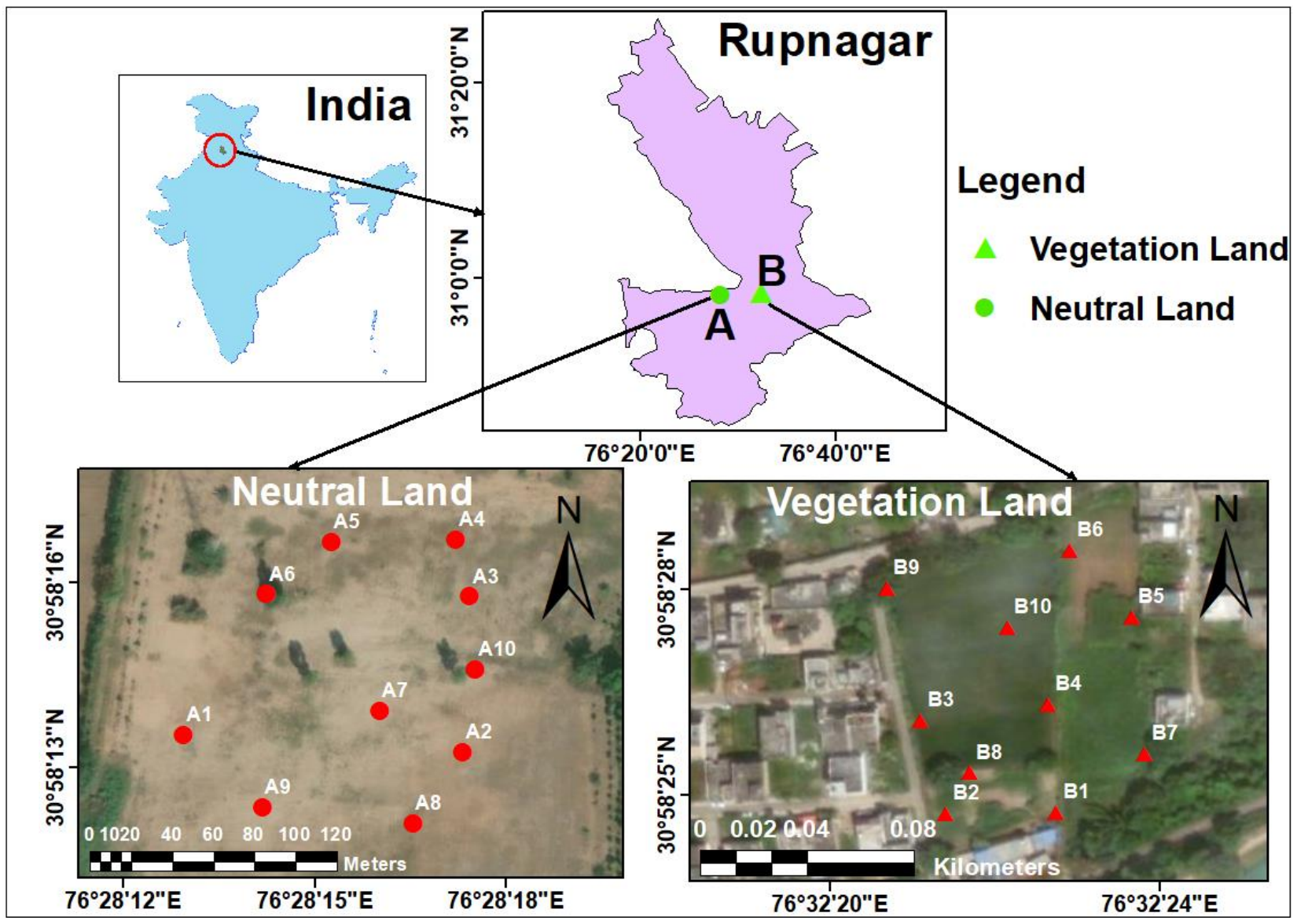

Figure 1. Distribution of soil sampling locations across both neutral and vegetation land within the study area.

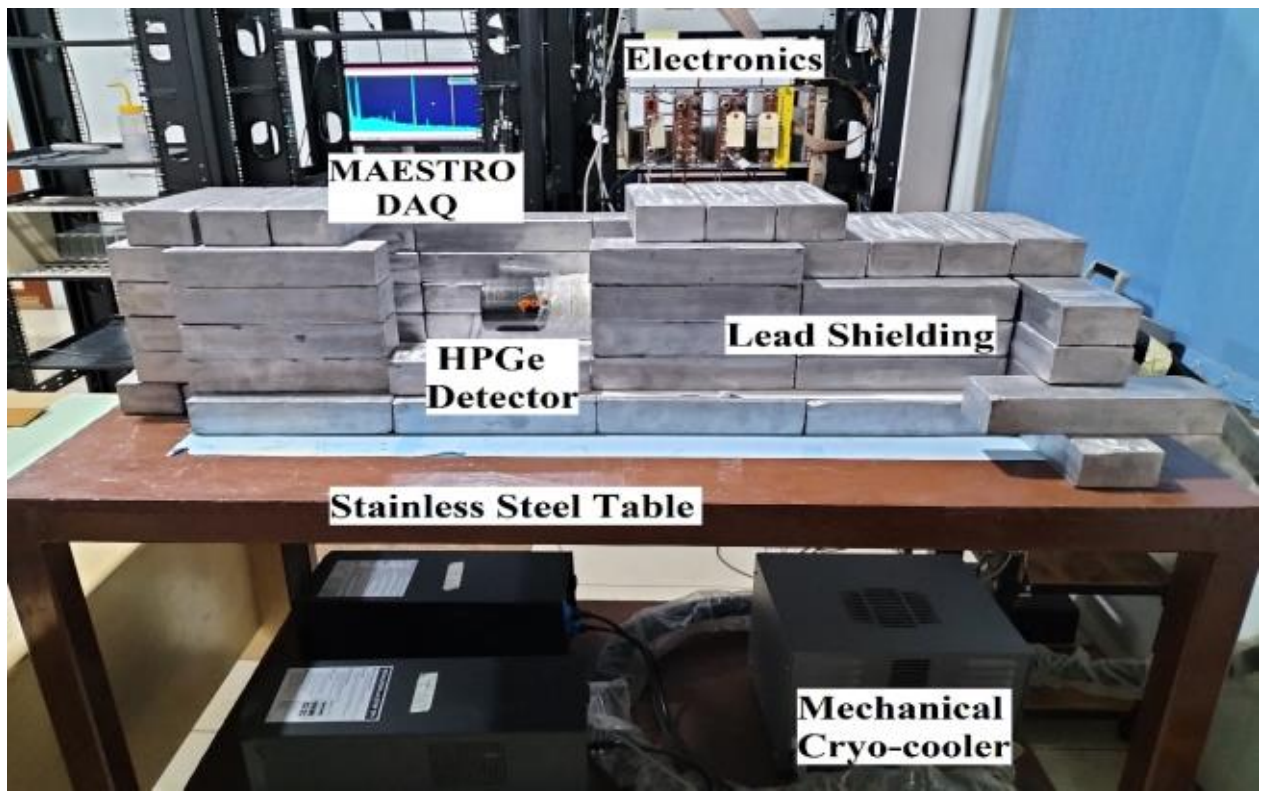

Figure 2(a). The lateral view of the experimental setup comprises of gamma spectrometer (IIT Ropar Low Background Measurement setup)

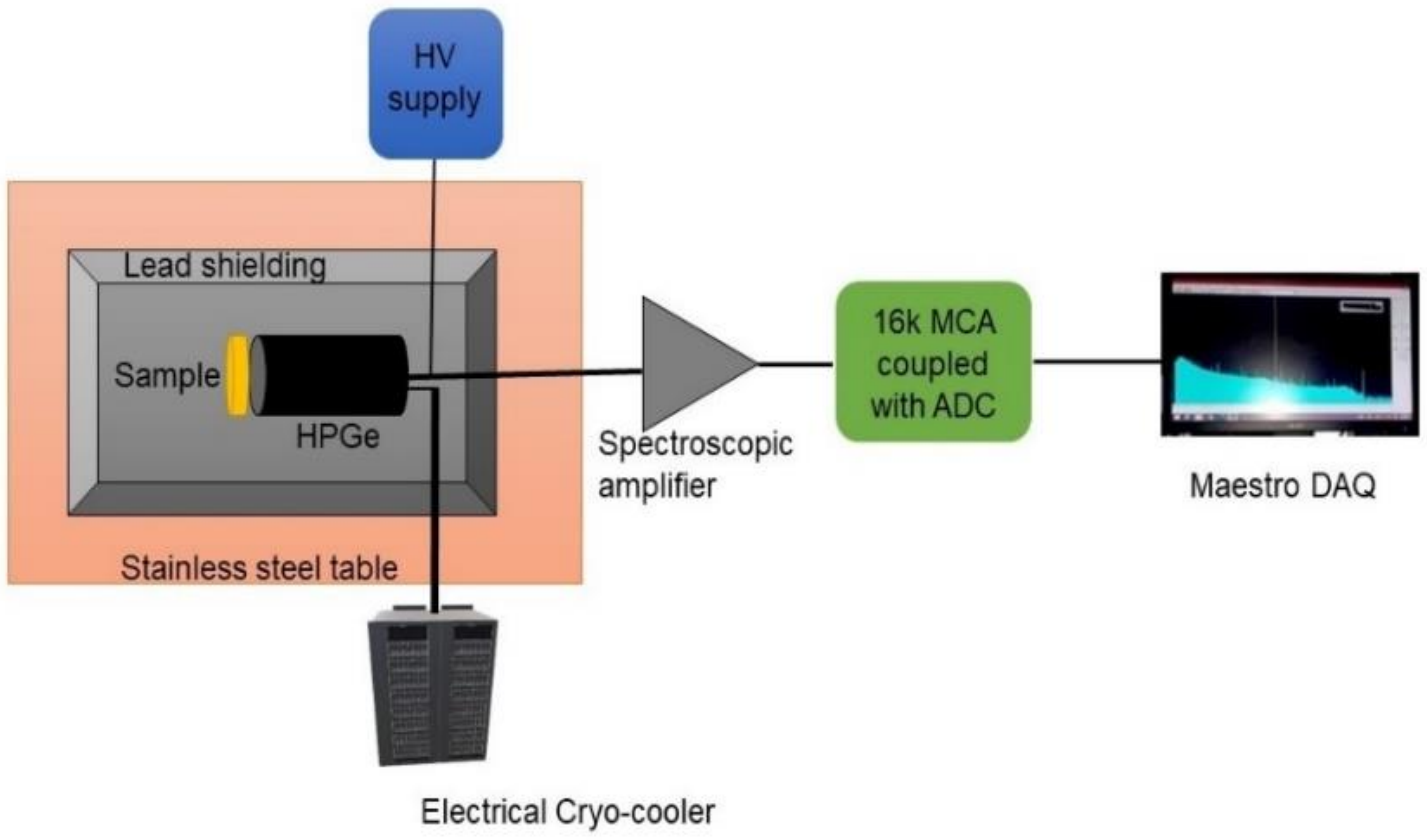

Figure 2(b). Schematics layout of the experimental setup

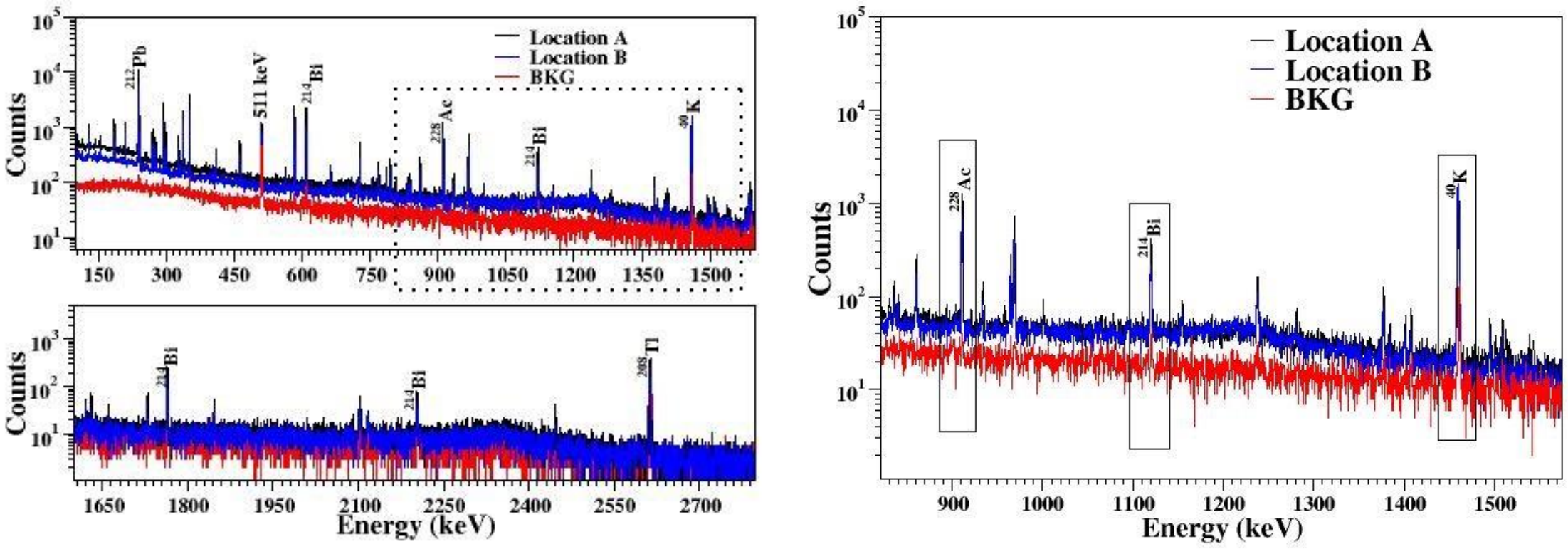

Figure 3. Typical gamma spectra of soil samples associated with locations A and B compared with the ambient background spectra for a duration of t = 24 hours

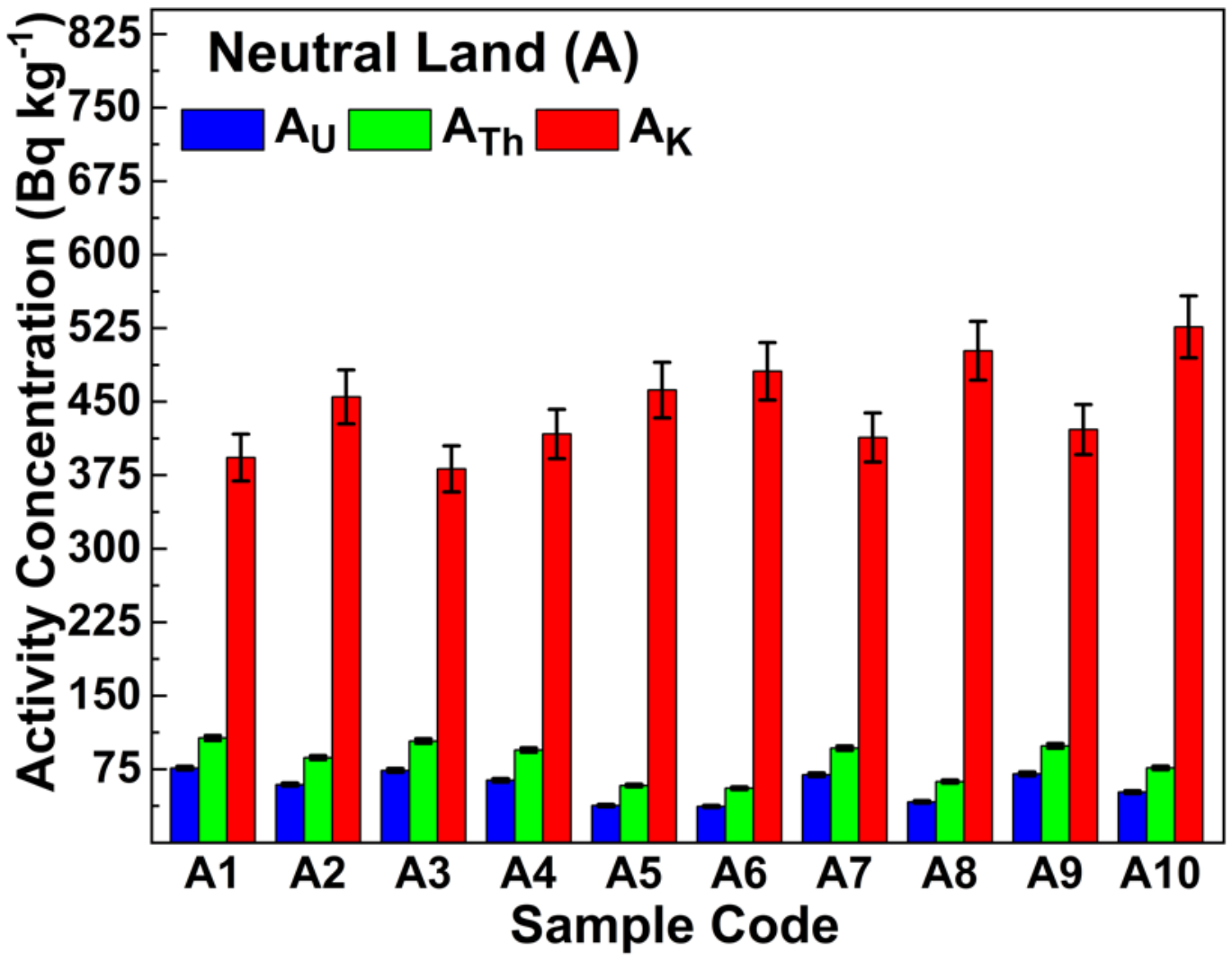

Figure 4. (a) Distribution of specific radioactivity of $^{238}$U, $^{232}$Th, and $^{40}$K in soil samples collected from neutral land (A)

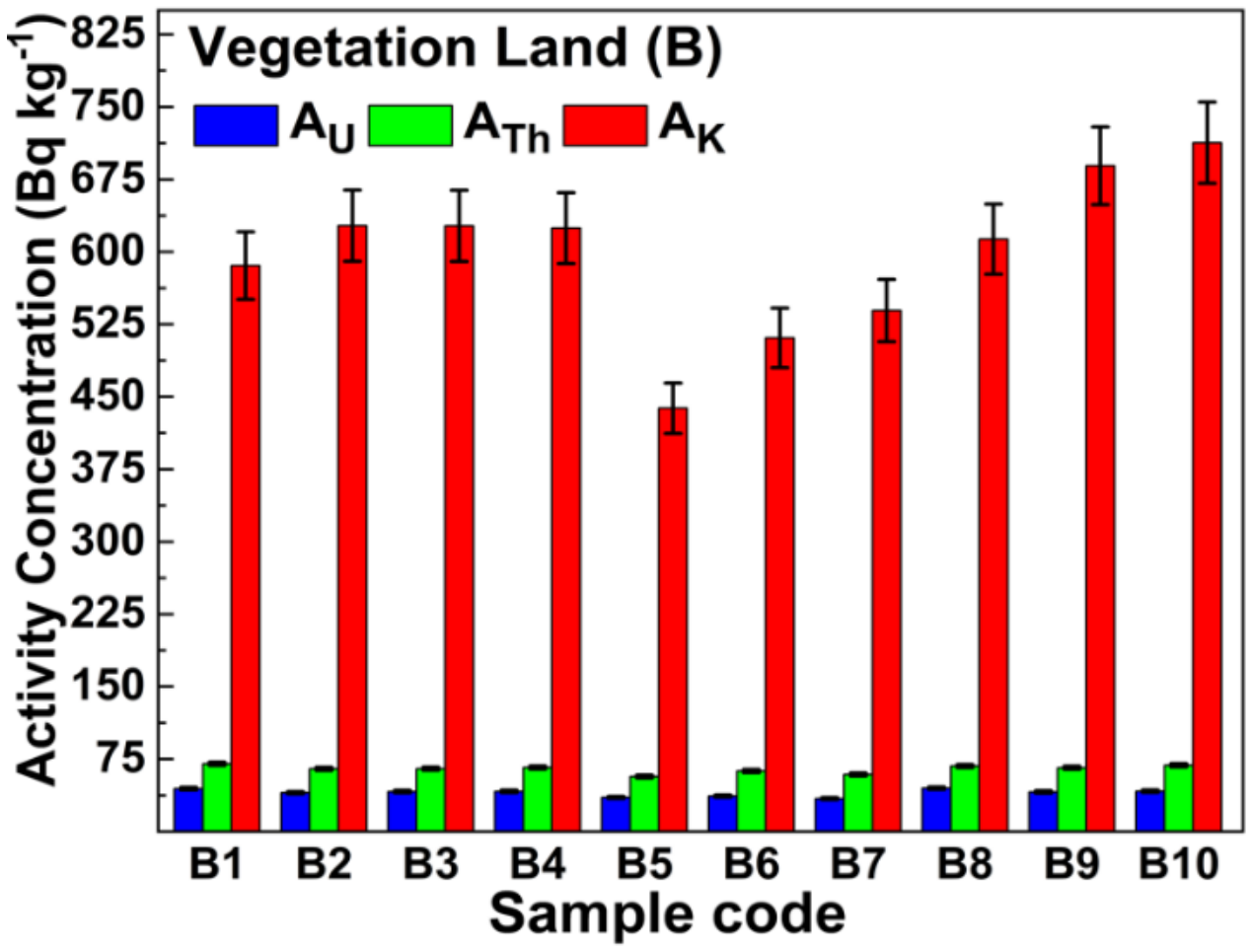

Figure 4. (b) Distribution of specific radioactivity of $^{238}$U, $^{232}$Th, and $^{40}$K in soil samples collected from vegetation land (B)

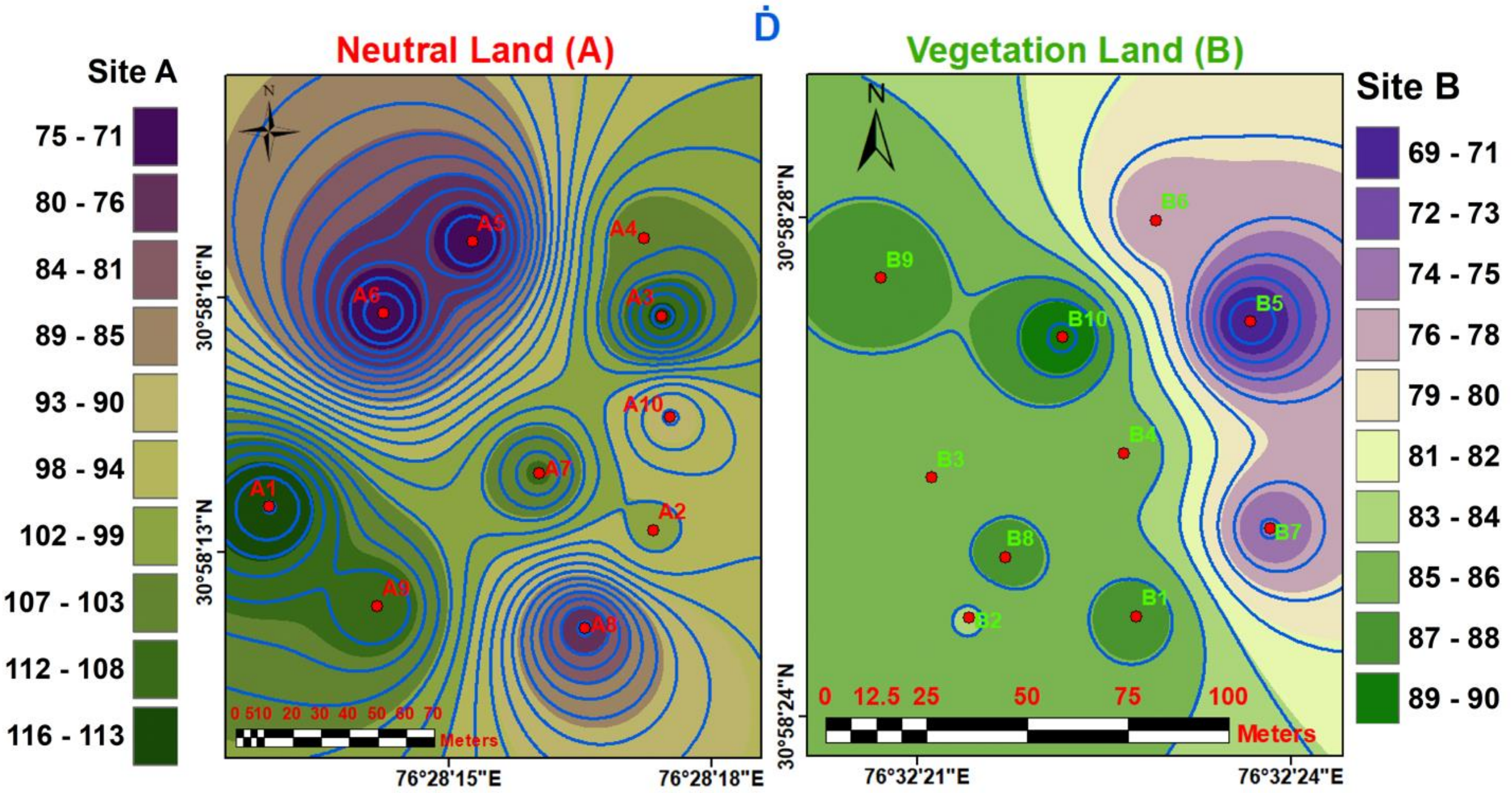

Figure 5. Visualization of the distribution pattern of air absorbed dose rate (Ḋ) in contour map with various contour lines.

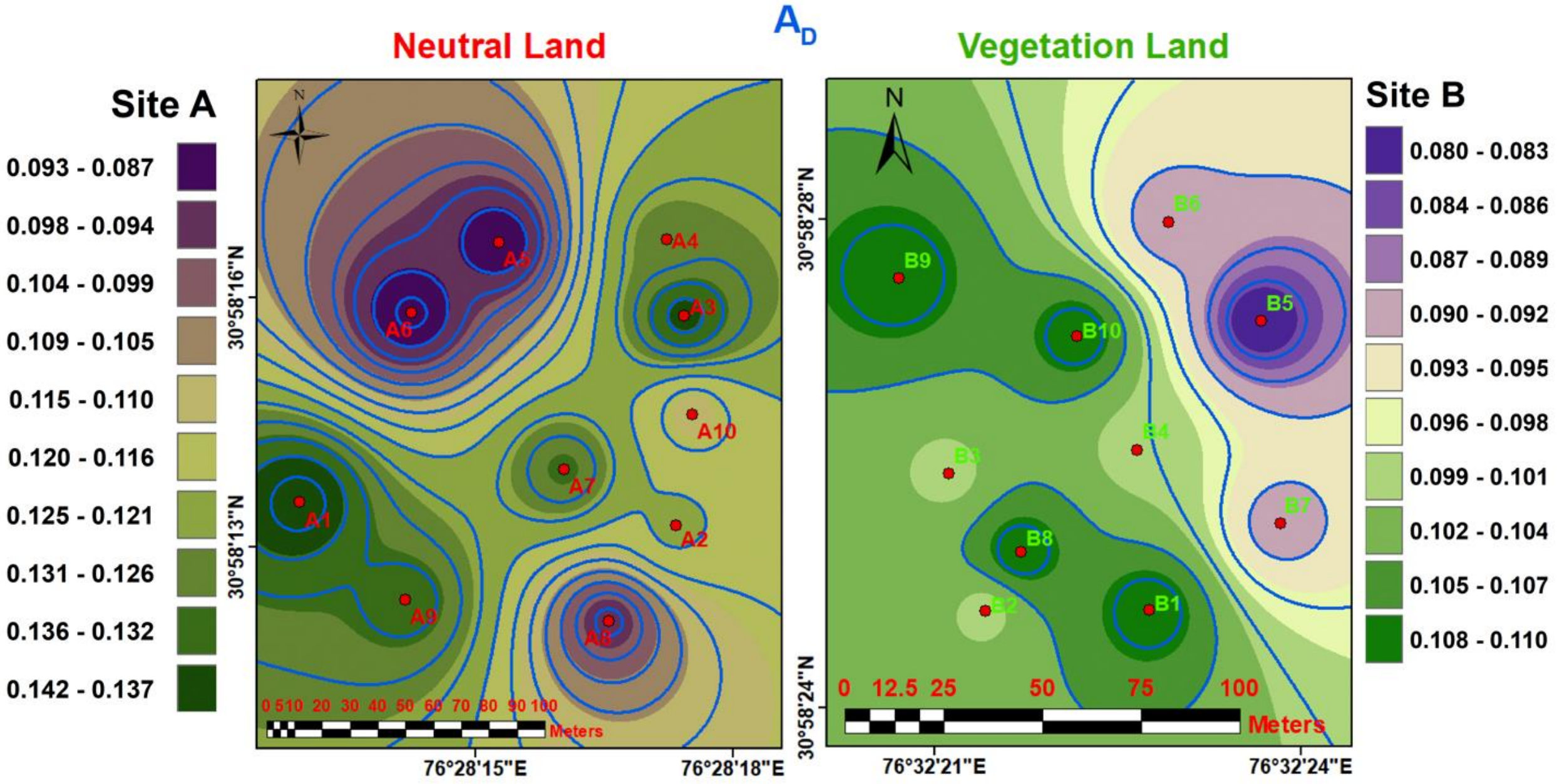

Figure 6. Visualization of the distribution pattern of annual effective dose ($A_D$) in contour map with various contour lines.

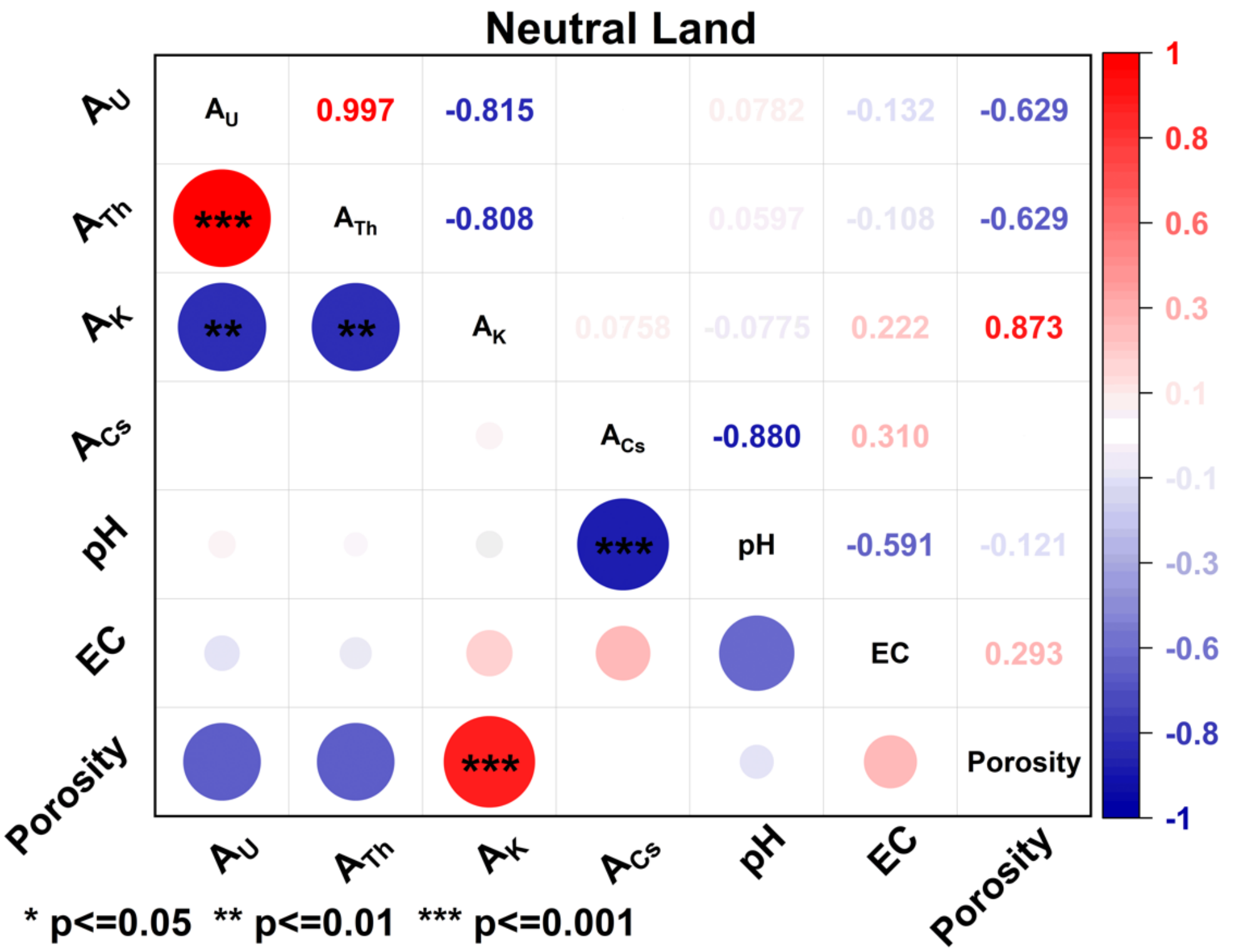

Figure 7 (a). Correlation between the studied radionuclides and physicochemical properties in the soil of neutral land

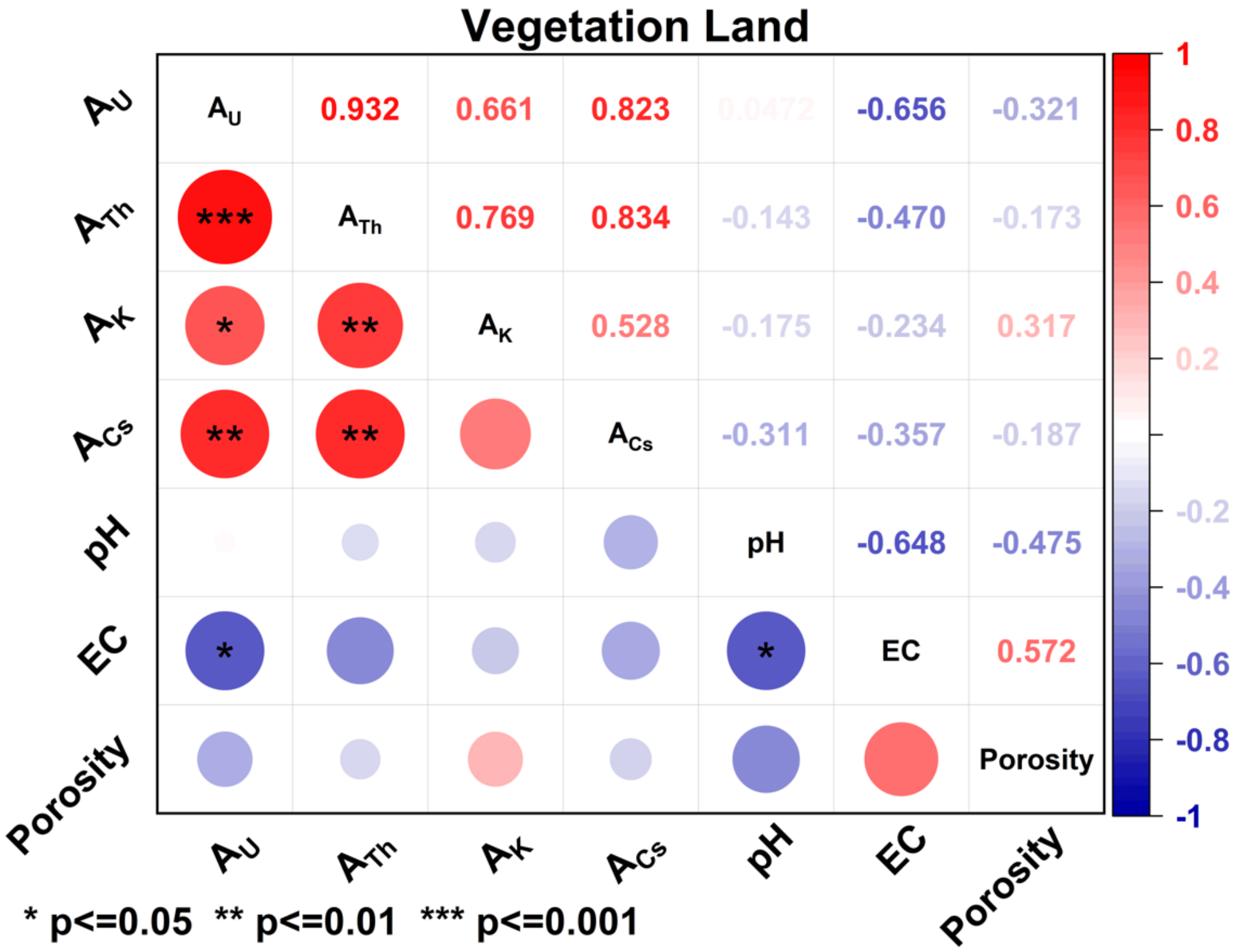

Figure 7 (b). Correlation between the studied radionuclides and physicochemical properties in the soil of vegetation land

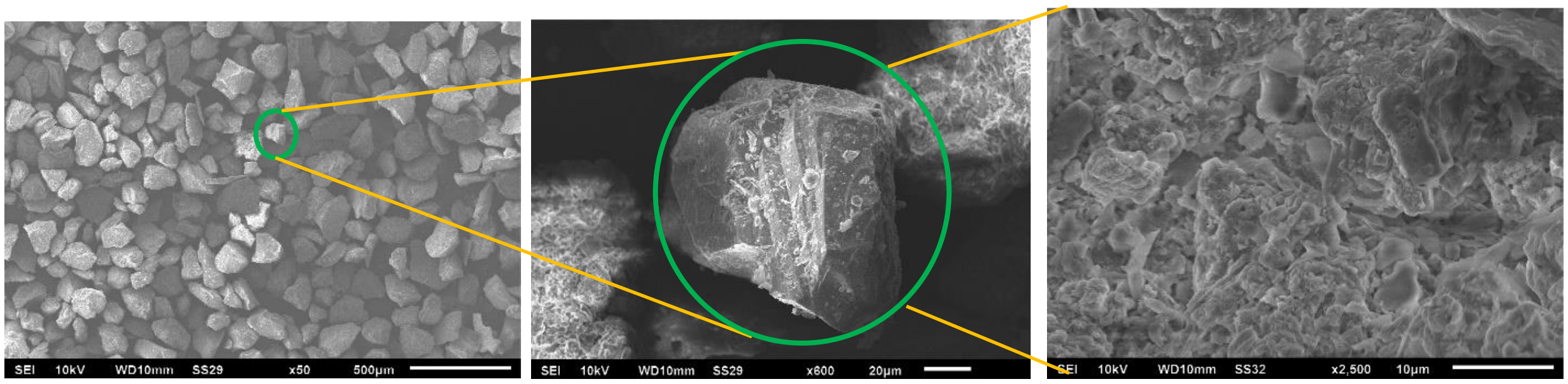

(a) (b) (c)

Figure 8. SEM micrographs of soil sample of neutral land having magnification: (a)- 50 times, (b)- 600 times, and (c)- 2500 times

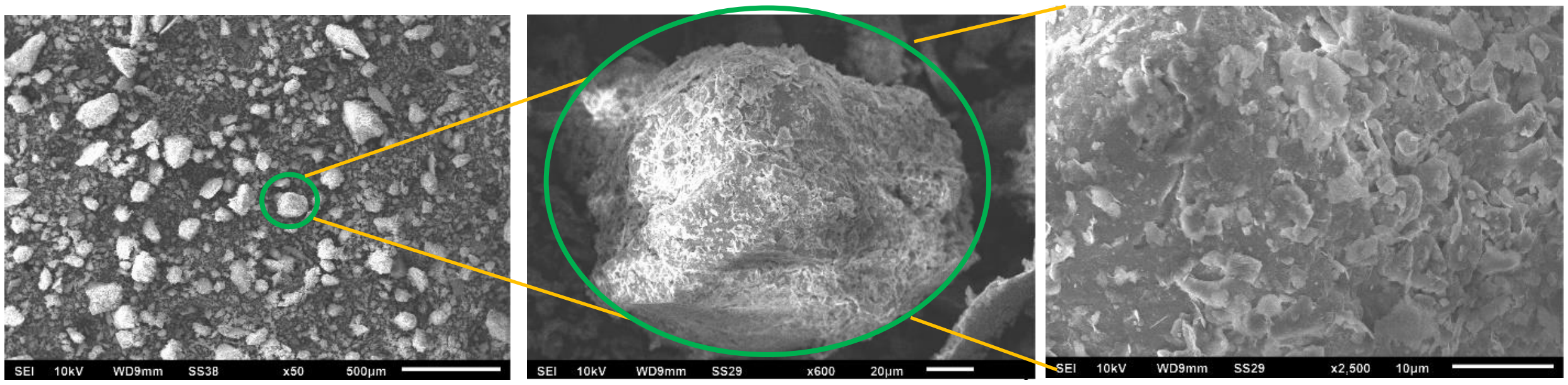

(a) (b) (c)

Figure 9. SEM images of soil samples of vegetation land having magnification: (a)- 50 times, (b)- 600 times, and (c)- 2500 times

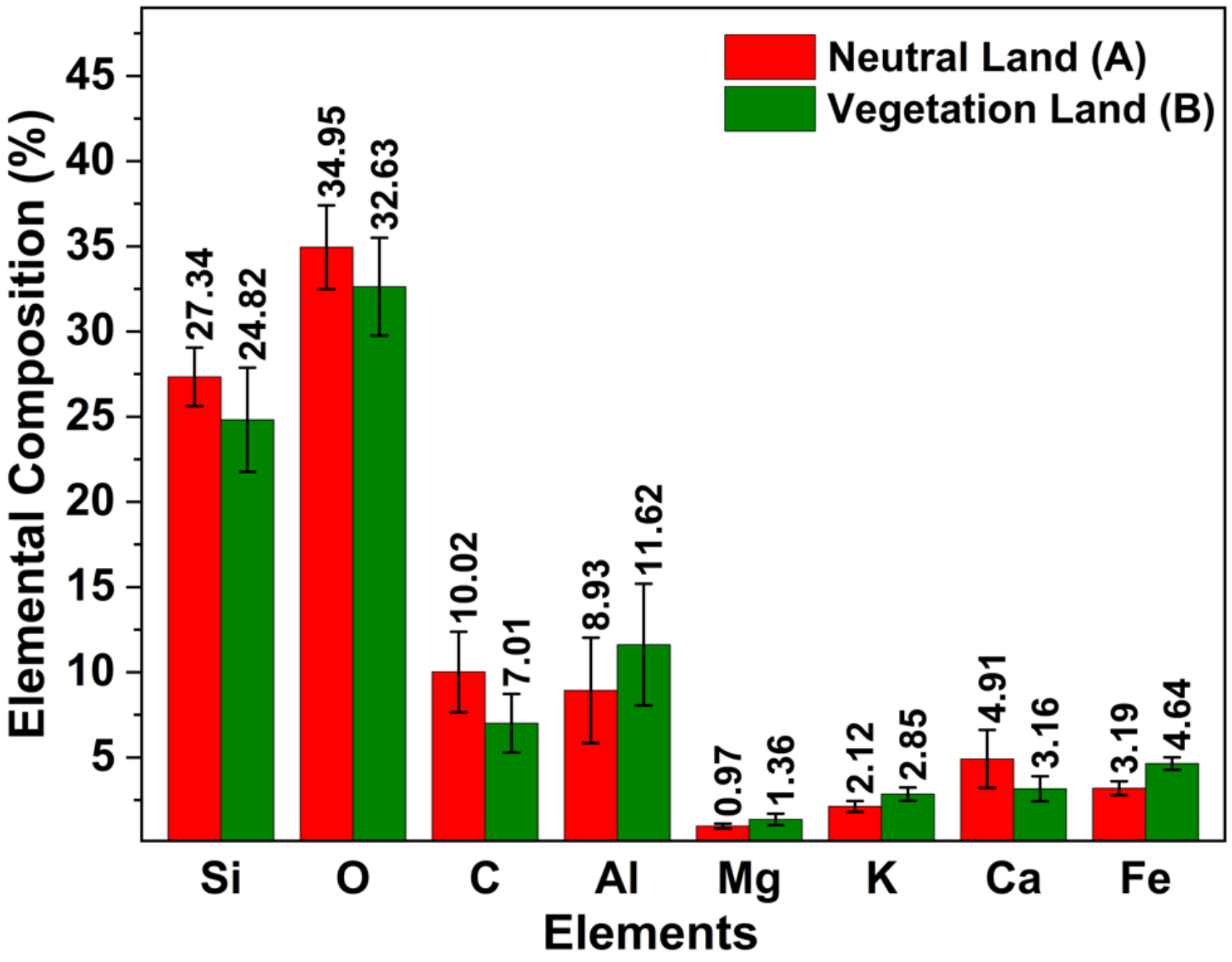

Figure 10. Elemental composition of major and minor elements detected using EDX spectroscopy

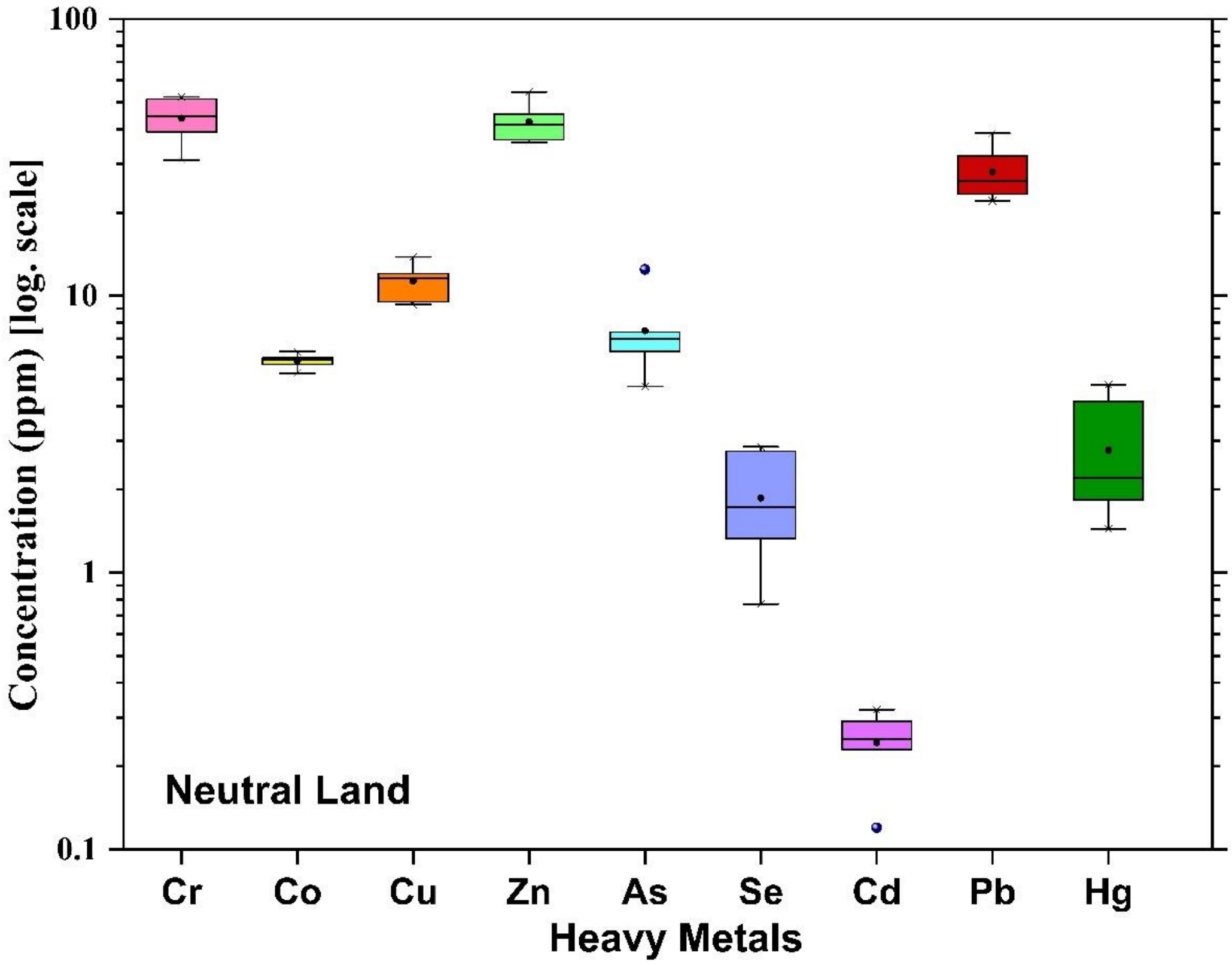

Figure 11 (a). Heavy metal concentration distribution in soil samples collected from neutral land

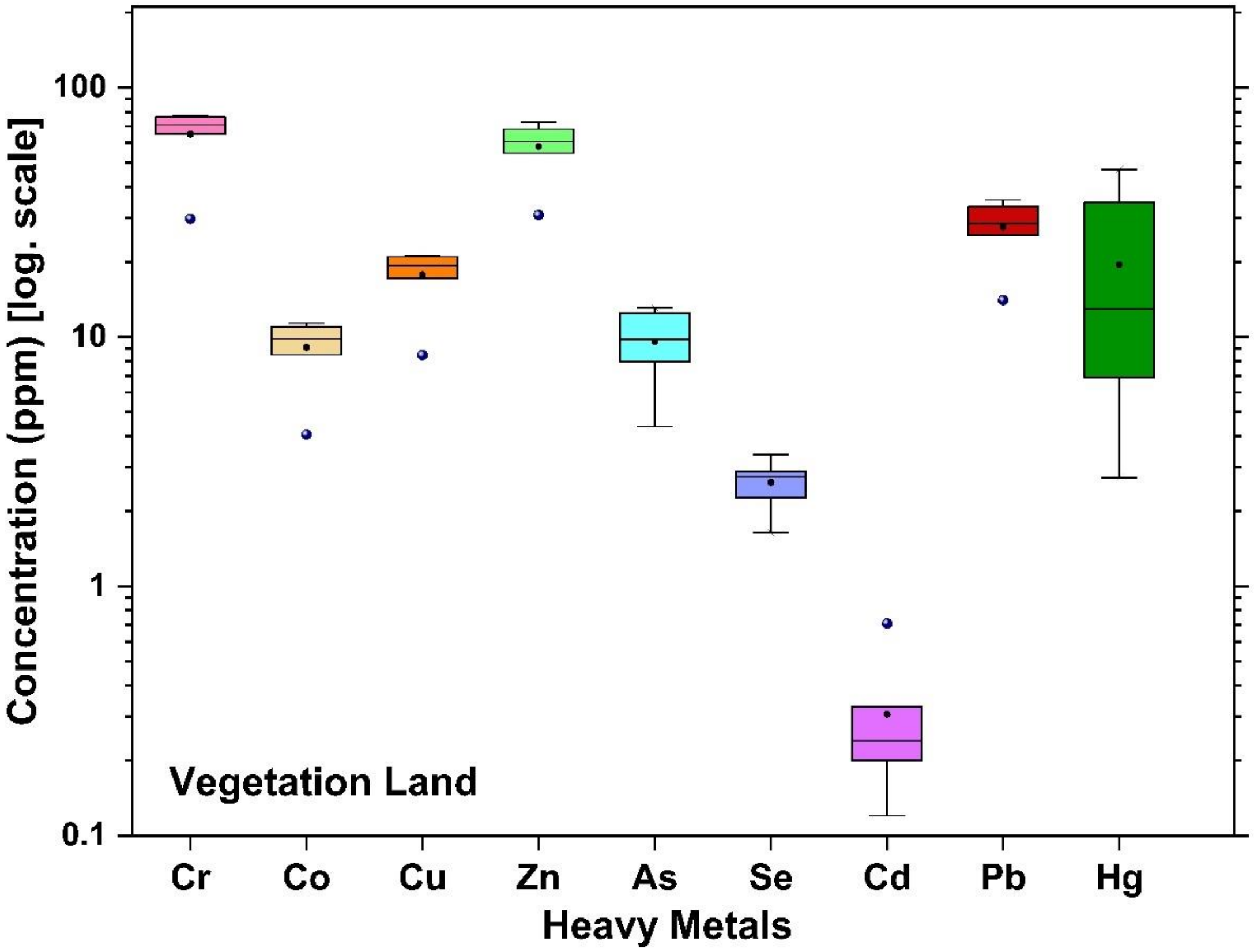

Figure 11 (b). Heavy metal concentration distribution in soil samples collected from vegetation land

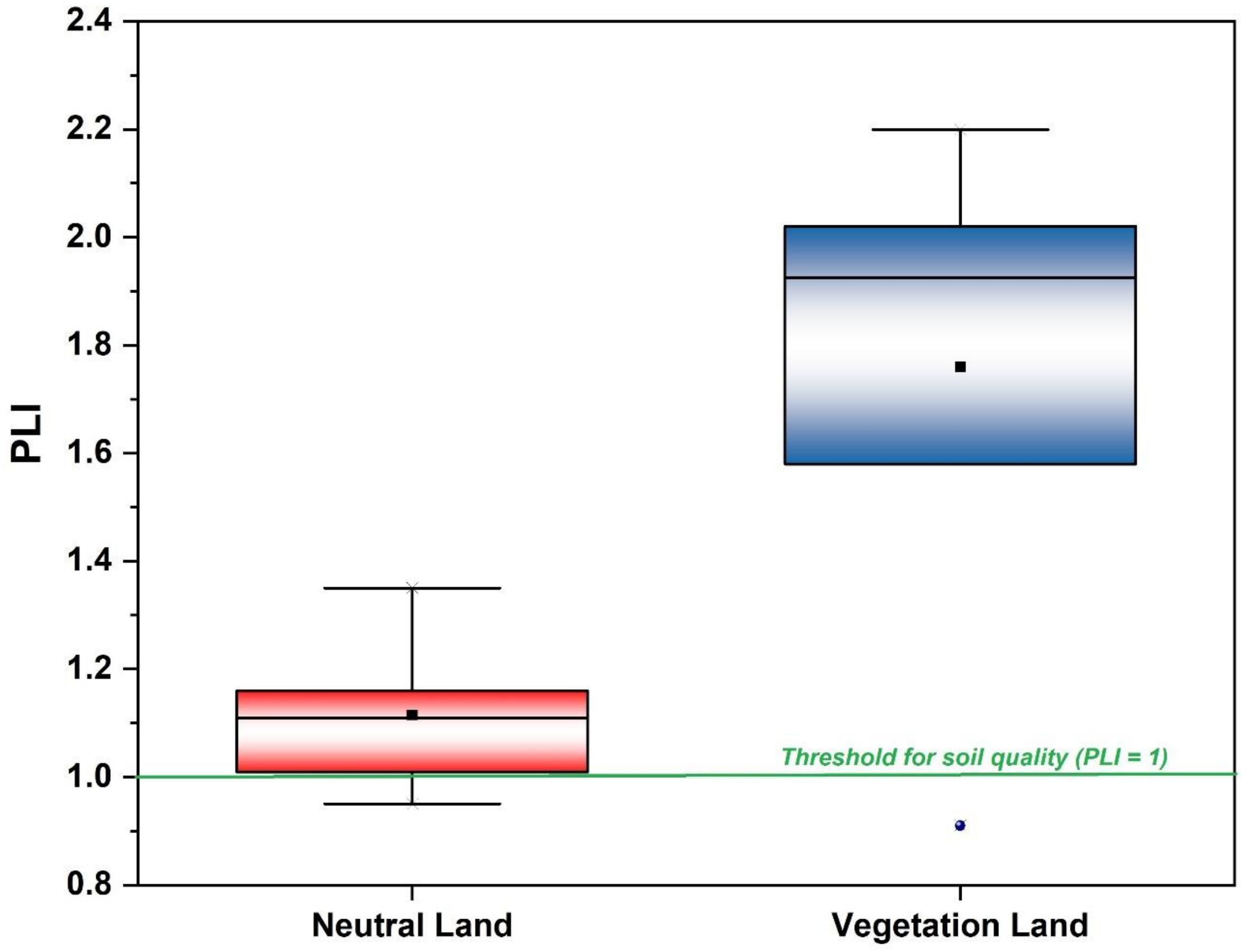

Figure 12. Pollution loaded index in neutral and vegetation Land

**Table 1. Energy and transition probability for various radionuclides of interest**

| Parents Radionuclides | Daughter Radionuclides | Gamma energy, $E_\gamma$ (keV) | Transition probability, P (%) |
|---|---|---|---|
| $^{238}U$ | $^{214}Pb$ | 295.2 | 18.47 |
| | $^{214}Pb$ | 351.9 | 35.60 |
| | $^{214}Bi$ | 609.3 | 45.49 |
| | $^{214}Bi$ | 1120.3 | 14.92 |
| | $^{214}Bi$ | 1764.5 | 15.30 |
| $^{232}Th$ | $^{212}Pb$ | 238.6 | 43.60 |
| | $^{228}Ac$ | 338.3 | 11.27 |
| | $^{208}Tl$ | 583.2 | 85.00 |
| | $^{228}Ac$ | 911.2 | 25.80 |
| | $^{228}Ac$ | 969.0 | 15.80 |
| | $^{208}Tl$ | 2614.5 | 99.80 |
| $^{40}K$ | $^{40}Ar$ | 1460.8 | 10.66 |
| $^{137}Cs$ | $^{137}Ba$ | 661.67 | 85.10 |

**Table 2. Activity concentration of various radionuclides and the associated radiological parameters at different locations in neutral (A) and vegetation land (B)**

| **Sample code** | **$A_U$ (Bq kg$^{-1}$)** | **$A_{Th}$ (Bq kg$^{-1}$)** | **$A_K$ (Bq kg$^{-1}$)** | **$A_{U-235}$ (Bq kg$^{-1}$)** | **$A_{Cs}$ (Bq kg$^{-1}$)** | **$^{235}U/^{238}U$** | **$Ra_{eq}$ (Bq kg$^{-1}$)** | **Air absorbed dose rate, $\dot{D}$ (nGy h$^{-1}$)** | **Annual effective dose, $D_E$ (mSv)** |
|---|---|---|---|---|---|---|---|---|---|
| | | | | Neutral Land (A) | | | | | |
| A1 | 76.13 ± 1.96 | 106.82 ± 2.53 | 392.99 ± 23.93 | 3.15 ± 0.67 | 0.19 ± 0.08 | 0.041 | 258.96 | 116.08 | 0.14 |
| A2 | 59.26 ± 1.58 | 86.72 ± 2.08 | 454.78 ± 27.55 | 2.85 ± 0.54 | 6.33 ± 0.56 | 0.048 | 218.13 | 98.72 | 0.12 |
| A3 | 73.69 ± 1.98 | 103.70 ± 2.43 | 381.37 ± 23.49 | 3.34 ± 0.63 | BDL | 0.045 | 251.17 | 112.58 | 0.14 |
| A4 | 63.72 ± 1.79 | 94.52 ± 2.25 | 416.96 ± 25.17 | 2.91 ± 0.58 | 0.22 ± 0.11 | 0.046 | 230.82 | 103.92 | 0.13 |
| A5 | 37.88 ± 1.16 | 58.34 ± 1.43 | 461.74 ± 28.34 | 1.95 ± 0.41 | 0.49 ± 0.26 | 0.051 | 156.74 | 71.99 | 0.09 |
| A6 | 37.00 ± 1.17 | 55.58 ± 1.43 | 480.91 ± 29.30 | 2.23 ± 0.42 | 0.25 ± 0.05 | 0.060 | 153.39 | 70.72 | 0.09 |
| A7 | 69.34 ± 1.87 | 96.41 ± 2.25 | 413.41 ± 25.07 | 3.34 ± 0.64 | BDL | 0.048 | 238.87 | 107.51 | 0.13 |
| A8 | 41.59 ± 1.25 | 62.33 ± 1.58 | 501.86 ± 30.00 | 2.28 ± 0.43 | 0.22 ± 0.13 | 0.055 | 169.24 | 77.79 | 0.10 |
| A9 | 70.15 ± 1.94 | 98.78 ± 2.37 | 421.51 ± 25.53 | 3.37 ± 0.68 | 0.45 ± 0.30 | 0.048 | 243.69 | 109.65 | 0.13 |
| A10 | 51.54 ± 1.47 | 76.26 ± 1.85 | 526.26 ± 31.69 | 2.90 ± 0.54 | BDL | 0.056 | 200.96 | 91.82 | 0.11 |
| **Minimum** | **37.00** | **55.58** | **381.37** | **1.95** | **BDL** | **0.041** | **153.39** | **70.72** | **0.09** |
| **Maximum** | **76.13** | **106.82** | **526.26** | **3.37** | **6.33** | **0.060** | **258.96** | **116.08** | **0.14** |
| **Average** | **58.03** | **83.95** | **445.18** | **2.83** | **1.16** | **0.050** | **212.20** | **96.08** | **0.12** |
| **SD** | **14.29** | **18.43** | **45.38** | **0.49** | **2.11** | **0.005** | **37.81** | **16.22** | **0.02** |
| Vegetation Land (B) | | | | | | | | | |
| B1 | 44.35 ± 1.34 | 70.04 ± 1.75 | 585.80 ± 34.96 | 2.60 ± 0.48 | 4.07 ± 0.42 | 0.059 | 189.47 | 87.22 | 0.11 |
| B2 | 40.13 ± 1.20 | 64.85 ± 1.63 | 627.00 ± 36.82 | 2.03 ± 0.38 | BDL | 0.051 | 181.00 | 83.86 | 0.10 |
| B3 | 41.29 ± 1.25 | 64.98 ± 1.65 | 626.82 ± 36.93 | 2.40 ± 0.47 | 2.49 ± 0.33 | 0.058 | 182.34 | 84.46 | 0.10 |
| B4 | 41.49 ± 1.25 | 66.23 ± 1.63 | 624.50 ± 36.66 | 2.28 ± 0.42 | 1.92 ± 0.33 | 0.055 | 184.14 | 85.21 | 0.10 |
| B5 | 35.27 ± 1.07 | 56.94 ± 1.49 | 438.11 ± 26.00 | 1.92 ± 0.39 | 0.23 ± 0.12 | 0.054 | 150.31 | 68.96 | 0.08 |
| B6 | 36.71 ± 1.18 | 62.57 ± 1.57 | 510.92 ± 30.65 | 2.10 ± 0.43 | 0.58 ± 0.21 | 0.057 | 165.40 | 76.06 | 0.09 |
| B7 | 34.06 ± 1.03 | 59.00 ± 1.49 | 539.22 ± 32.30 | 2.06 ± 0.40 | 0.20 ± 0.17 | 0.060 | 159.82 | 73.86 | 0.09 |
| B8 | 44.79 ± 1.28 | 67.70 ± 1.68 | 613.05 ± 36.35 | 2.42 ± 0.48 | 2.71 ± 0.42 | 0.054 | 188.66 | 87.15 | 0.11 |
| B9 | 40.92 ± 1.20 | 65.90 ± 1.62 | 689.06 ± 40.32 | 2.36 ± 0.45 | 1.56 ± 0.31 | 0.058 | 188.07 | 87.44 | 0.11 |
| B10 | 41.68 ± 1.29 | 68.61 ± 1.70 | 712.90 ± 42.02 | 2.41 ± 0.49 | 3.32 ± 0.31 | 0.058 | 194.53 | 90.42 | 0.11 |
| **Minimum** | **34.06** | **56.94** | **438.11** | **1.92** | **BDL** | **0.051** | **150.31** | **68.96** | **0.08** |
| **Maximum** | **44.79** | **70.04** | **712.90** | **2.60** | **4.07** | **0.060** | **194.53** | **90.42** | **0.11** |
| **Average** | **40.07** | **64.68** | **596.74** | **2.26** | **1.90** | **0.056** | **178.37** | **82.46** | **0.10** |
| **SD** | **3.43** | **3.92** | **77.99** | **0.21** | **1.30** | **0.003** | **13.93** | **6.66** | **0.01** |
| **World average (UNSCEAR 2008)** | **33** | **45** | **420** | **-** | **51** | **-** | **129.59** | **58** | **0.07** |

*BDL: Below Detection Limit

**Table 3. Physio-chemical properties of soil at different locations in both neutral (A) and vegetation land (B)**

| **Sample code** | **pH** | **Electrical Conductivity (µS cm$^{-1}$)** | **Porosity (%)** | **Sample code** | **pH** | **Electrical Conductivity (µS cm$^{-1}$)** | **Porosity (%)** |
|---|---|---|---|---|---|---|---|
| A1 | 8.80 | 74.1 | 37.50 | B1 | 8.68 | 73.70 | 40.00 |
| A2 | 8.16 | 66.80 | 40.00 | B2 | 8.85 | 85.10 | 42.50 |
| A3 | 8.86 | 48.27 | 37.50 | B3 | 8.66 | 104.67 | 42.50 |
| A4 | 8.9 | 59.70 | 37.50 | B4 | 8.81 | 87.60 | 42.50 |
| A5 | 8.93 | 54.43 | 40.00 | B5 | 8.74 | 115.40 | 45.00 |
| A6 | 8.64 | 66.77 | 42.50 | B6 | 8.48 | 145.30 | 45.00 |
| A7 | 8.99 | 46.26 | 37.50 | B7 | 8.57 | 115.57 | 47.50 |
| A8 | 9.01 | 52.50 | 40.00 | B8 | 8.59 | 77.20 | 42.50 |
| A9 | 9.07 | 46.70 | 40.00 | B9 | 8.67 | 104.30 | 52.38 |
| A10 | 8.90 | 64.60 | 45.00 | B10 | 8.34 | 130.33 | 50.00 |
| **Minimum** | **8.16** | **46.26** | **37.50** | **Minimum** | **8.34** | **73.73** | **40.00** |
| **Maximum** | **9.07** | **74.17** | **45.00** | **Maximum** | **8.85** | **145.30** | **52.38** |
| **Average** | **8.83** | **58.02** | **39.75** | **Average** | **8.64** | **103.92** | **44.99** |
| **SD** | **0.26** | **9.80** | **2.49** | **SD** | **0.15** | **23.39** | **3.88** |

**Table 4. Heavy metal concentration for both neutral (A) and vegetation land (B) with background and permissible limits (in ppm)**

| | Heavy Metals | Cr | Co | Cu | Zn | As | Se | Cd | Pb | Hg |
|---|---|---|---|---|---|---|---|---|---|---|
| **Neutral Land 'A'** | Range | 31.00 - 52.39 | 5.26 - 6.29 | 9.31- 13.82 | 35.85 - 54.47 | 4.71 - 12.47 | 0.77 - 2.85 | 0.12 - 0.32 | 2.20 - 38.69 | 1.44 - 4.78 |
| | Avg. ± SD | 43.84 ± 7.41 | 5.83 ± 0.32 | 11.32 ± 1.54 | 42.56 ± 6.23 | 7.48 ± 2.4 | 1.86 ± 0.74 | 0.24 ± 0.06 | 28.02 ± 5.73 | 2.77 ± 1.25 |
| **Vegetation Land 'B'** | Range | 29.77 - 76.93 | 4.06 - 11.35 | 8.47 - 21.16 | 30.86 - 72.89 | 4.37 – 13.08 | 1.64 - 3.38 | 0.12 - 0.71 | 14.04 - 35.55 | 6.87 - 46.97 |
| | Avg. ± SD | 65.02 ± 16.39 | 9.08 ± 2.48 | 17.74 ± 4.4 | 58.00 ± 13.65 | 9.58 ± 3.08 | 2.61 ± 0.55 | 0.31 ± 0.19 | 27.58 ± 6.89 | 19.51 ± 15.94 |
| **Bkg*** | Kuhad et al. 1989 and Kumar et al 2019 | 114 | 15.2 | 56.5 | 22.1 | - | - | - | 13.1 | - |
| **Permissible limits** | Awashthi et al. 2000 | - | 300 - 600 | 135 - 270 | 300 - 600 | - | - | 3.00 – 6.00 | 250 - 500 | - |
| | USEPA 2002 | 11 | - | 270 | 1100 | - | - | 0.48 | 200 | 1 |
| | WHO/FAO 2007 | 100 | 50 | 100 | 50 | 20 | - | 0.30 | 20 | 0.3 |

*Bkg is the Indian natural background for soil

**Table 5. Classification of soil based on various pollution indices**

| **CF** (Hakanson 1980) | | **PLI** (Tomlinson et al. 1980) | | $\mathbf{I_{geo}}$ (Muller 1969) | | $\mathbf{E_i}$ (Hakanson et al. 1980) | | **RI** (Hakanson et al. 1980) | |
|---|---|---|---|---|---|---|---|---|---|
| **CF** < 1 | Low contamination | **PLI** < 1 | Unpolluted soil | $\mathbf{I_{geo}} \leq 0$ | Uncontaminated | $\mathbf{E_i}$ < 40 | low | < 150 | low |
| 1 < **CF**< 3 | Moderate contamination | **PLI** = 1 | Pollutant levels present | $0 < \mathbf{I_{geo}} < 1$ | Uncontaminated to moderately contaminated | 40 < $\mathbf{E_i}$ < 80 | Fairly | 150-300 | Moderate |
| 3 < **CF**< 6 | Considerable contamination | **PLI** > 1 | Deterioration of soil quality | $1 < \mathbf{I_{geo}} < 2$ | Moderately contaminated | 80-160 | Considerable | 300-600 | High |
| > 6 | Very high contamination | | | $2 < \mathbf{I_{geo}} < 3$ | Moderately to strongly contaminated | 160-320 | High | > 600 | Very high |
| | | | | $3 < \mathbf{I_{geo}} < 4$ | Strongly contaminated | > 320 | Very high | | |
| | | | | $4 < \mathbf{I_{geo}} < 5$ | Strongly to Extremely contaminated | | | | |
| | | | | $\mathbf{I_{geo}} \geq 5$ | Extremely contaminated | | | | |

**Table 6. Statistical summary of pollution indices due to various heavy metals observed in neutral and vegetation land**

| Location | Pollution index | Statistical parameter | HEAVY METAL | | | | | | | | |
|---|---|---|---|---|---|---|---|---|---|---|---|
| | | | Cr | Co | Cu | Zn | As | Se | Cd | Pb | Hg |
| Neutral Land | CF | Range | 0.89 - 1.50 | 0.53 - 0.63 | 0.37 - 0.55 | 0.50 - 0.77 | 3.14 - 8.32 | 0.02 - 0.06 | 1.23 - 3.21 | 1.10 - 1.93 | 14.35 - 47.76 |
| | | Avg. ± SD | 1.25 ± 0.21 | 0.58 ± 0.03 | 0.45 ± 0.06 | 0.60 ± 0.09 | **4.99 ± 1.60** | 0.04 ± 0.01 | 2.48 ± 0.64 | 1.40 ± 0.29 | **27.68 ± 12.45** |
| | $I_{geo}$ | Range | 0.18 - 0.30 | 0.11- 0.12 | 0.07 - 0.11 | 0.10 - 0.15 | 0.63 - 1.67 | 0.00 - 0.01 | 0.25 - 0.65 | 0.22 - 0.39 | 2.88 - 9.59 |
| | | Avg. ± SD | 0.25 ± 0.04 | 0.12 ± 0.01 | 0.09 ± 0.01 | 0.12 ± 0.02 | **1.00 ± 0.32** | 0.007 ± 0.003 | 0.49 ± 0.13 | 0.28 ± 0.06 | **5.55 ± 2.49** |
| | $E_i$ | Range | 1.77 - 2.99 | - | 1.86 - 2.76 | 0.50 - 0.77 | 31.40 - 83.15 | - | 37.04 - 96.43 | 5.51 - 9.67 | 574.00 – 1910.40 |
| | | Avg. ± SD | 2.51 ± 0.42 | - | 2.26 ± 0.31 | 0.60 ± 0.09 | 49.88 ± 16.03 | - | 74.29 ± 19.08 | 7.00 ± 1.43 | **1107.00 ± 498.10** |
| Vegetation Land | CF | Range | 0.85 - 2.20 | 0.41 - 1.13 | 0.34 - 0.85 | 0.43 - 1.03 | 2.91 - 8.72 | 0.03 - 0.07 | 1.19 - 7.24 | 0.70 - 1.78 | 27.32 – 469.66 |
| | | Avg. ± SD | 1.86 ± 0.47 | 0.91 ± 0.25 | 0.71 ± 0.18 | 0.82 ± 0.19 | **6.38 ± 2.05** | 0.05 ± 0.01 | **3.13 ± 1.95** | 1.38 ± 0.34 | **195.08 ± 159.36** |
| | $I_{geo}$ | Range | 0.17 - 0.44 | 0.08 - 0.23 | 0.06 - 0.17 | 0.08 - 0.21 | 0.58 - 1.75 | 0.00 - 0.01 | 0.24 - 1.45 | 0.14 - 0.36 | 5.48 – 94.25 |
| | | Avg. ± SD | 0.37 ± 0.09 | 0.18 ± 0.05 | 0.14 ± 0.03 | 0.16 ± 0.04 | **1.28 ± 0.41** | 0.010 ± 0.002 | 0.63 ± 0.39 | 0.28 ± 0.07 | **39.15 ± 31.99** |
| | $E_i$ | Range | 1.70 - 4.40 | - | 1.69 - 4.23 | 0.43 - 1.03 | 29.14 - 87.21 | - | 35.82 - 217.35 | 3.51 - 8.89 | 1092.80 – 18786.40 |
| | | Avg. ± SD | 3.72 ± 0.94 | - | 3.55 ± 0.88 | 0.82 ± 0.19 | 63.85 ± 20.53 | - | 93.88 ± 58.48 | 6.89 ± 1.72 | **7803.33 ± 6375.04** |